%% file: Environmetrics_submission_earlyJune2025/VaNBayes_environmetrics_final_version.tex
\documentclass[APA,STIX1COL]{WileyNJD-v2}

\usepackage{amsmath}
\usepackage{graphicx,psfrag,epsf}
\usepackage{enumerate}

\RequirePackage{amsthm,amsmath,amsfonts,amssymb}
\usepackage{fullpage,setspace}
\usepackage{graphicx}
\usepackage{bm} 
\usepackage{bbm}
\usepackage{lineno}
\usepackage{colortbl}
\usepackage{comment}
\usepackage{multirow}
\usepackage{algorithm}
\usepackage{algpseudocode}
\usepackage{makecell}
\usepackage{url}

\input{commands}

\articletype{Article Type}%

\received{}
\revised{}
\accepted{}

\begin{document}

\title{\color{black}Demonstrating the power and flexibility of variational assumptions for amortized neural posterior estimation in environmental applications}

\author[1]{Elliot Maceda*}

\author[2]{Emily C Hector}

\author[3]{Amanda Lenzi}

\author[1]{Brian J Reich}


\address[1]{\orgname{Department of Statistics, North Carolina State University}, \orgaddress{\state{Raleigh, North Carolina}, \country{USA}}}

\address[2]{\orgname{\color{black}Department of Biostatistics, University of Michigan}, \orgaddress{\state{Ann Arbor, Michigan}, \country{USA}}}

\address[3]{\orgname{School of Mathematics, University of Edinburgh}, \orgaddress{\city{Edinburgh}, \country{UK}}}

\corres{*Corresponding author, \\ \email{elliot.maceda@gmail.com}}

\presentaddress{5109, SAS Hall, 2311 Stinson Dr, Raleigh, NC 27607}

\abstract[Abstract]{Classic Bayesian methods with complex environmental models are frequently infeasible due to an intractable likelihood. Simulation-based inference methods, such as {\color{black}neural posterior estimation}, calculate posteriors without accessing a likelihood function by leveraging the fact that data can be quickly simulated from the model, but converge slowly and/or poorly in high-dimensional settings. In this paper, we {\color{black}suggest that imposing strict variational assumptions on the form of the posterior can often combat these computational issues}. Posterior distributions of model parameters are efficiently obtained by assuming a parametric form for the posterior, parametrized by the machine learning model, which is trained with the simulated data as inputs and the associated parameters as outputs. We show theoretically that {\color{black}if the parametric family of the variational posterior is correct,} our posteriors converge to the true posteriors in Kullback-Leibler divergence. We also provide tools to help us identify if our parametric assumption is close to the true posterior, and modeling options if that is not the case. Comprehensive simulation studies using environmental models highlight our method's robustness and versatility. An analysis of the Zika virus in Brazil provides a thorough case study.}



\maketitle


\section{Introduction}\label{s:intro}

Computational tools for Bayesian posterior inference in complex models are at best slow and at worst intractable when the likelihood is difficult to evaluate.
Examples of such models include binary data on a spatial network, spatial epidemiological models and extreme value models for climate data.  A variety of solutions have been developed to overcome the intractability of the likelihood function in these models. Virtually all of these methods share one element in common: they leverage the fact that data can be quickly simulated at a variety of model parameter configurations. 

Recently, a new philosophy has emerged to address computational difficulties with intractable likelihoods. Since data can be quickly simulated from the complex model at various parameter value inputs, an emulator can be fit to predict parameters given the simulated data. Given the ease of simulation, a user can generate an arbitrarily large training set to improve predictive performance of the emulator. Therefore, the only requirement is a sufficiently powerful surrogate model that can learn the parameter values that generated the observed data computationally efficiently. We survey relevant literature in Section \ref{s:literat} below and highlight limitations of existing approaches.

In this paper, we {\color{black}suggest that targeting marginal posteriors of key parameters of interest and imposing a parametric or semiparametric assumption to the posterior dramatically reduces the complexity of the problem, allowing simpler architecture and fewer simulations to converge. This simplification makes simulation-based inference more accessible for environmental applications. Our primary contributions are to demonstrate via extensive empirical studies that this approach is a viable approach for challenging environmental applications, and to illustrate its practical application including selecting and validating variational approximations. }


\subsection{Related Work}\label{s:literat}

\paragraph{Classic simulation based inference}
The most popular method in this umbrella is arguably Approximate Bayesian Computation (ABC) \citep{fearnhead2012constructing, frazier2018asymptotic, sisson2018handbook}.
In the simplest form of rejection ABC, parameters are drawn from the prior and used to simulate data from the assumed model. The parameters are retained in the posterior sample if the simulated data are sufficiently close to the observed data. 
To reduce rejection rates to manageable levels, lower-dimensional features (or summary statistics) are used instead of raw data to measure closeness between simulated and observed data.
More advanced variants include Markov Chain Monte Carlo ABC \citep{marjoram2003markov} and Sequential Monte Carlo ABC \citep{sisson2007sequential, peters2012sequential}, which guide simulations based on previously accepted parameter values. Another approach is based on creating a model for the likelihood by estimating the distribution of simulated data with traditional non-parametric density estimation such as histograms or kernel density estimation \citep{diggle1984monte}.
To mitigate the curse of dimensionality inherent in standard ABC methods when the number of model parameters is large, \citet{li2017extending} proposed a copula approach that first estimates the bivariate posterior for each pair of parameters separately and then combines these estimates together to estimate the joint posterior. 
We refer to \citet{cranmer2020frontier} for a recent review on simulation-based inference.

\paragraph{Neural-network based point estimation}
Recent breakthroughs in deep learning, such as the integration of automatic differentiation and probabilistic programming into the simulation code, have led to a fast growing area of research in parameter estimation of statistical models using neural networks. 
\cite{gerber2021fast} trained convolutional neural networks (CNNs) to learn the mapping between data and parameters of spatial covariance functions in Gaussian processes and achieved similar accuracy yet significant computational efficiency when compared to the maximum likelihood estimator (MLE). \citet{lenzi2023neural} used CNNs to estimate the parameters of max-stable processes, whose likelihoods are well known to be intractable even with small datasets, and  showed improvements in computational time and accuracy over a composite likelihood method.
They proposed a modified bootstrap approach for uncertainty quantification of these estimators.
Recent advancements have successfully incorporated replicated data in estimation \citep{sainsbury2022fast}, neural networks for irregular spatial data \citep{sainsbury2023neural} and censoring information \citep{jordan2023neural}.

A drawback of these approaches is that the estimators will inevitably be biased towards the parameter region used to simulate the training data.
To avoid this issue, \citet{lenzi2023towards} proposed a sequential approach that modifies the training data using dynamically updated prior distributions by making use of the observed data. Nonetheless, inference remains challenging in high dimensions and large parameter spaces, and is typically restricted to summaries of the posterior distribution rather than the full posterior itself. Our proposed method {\color{black}goes beyond point estimation by} learning a parametric approximation to the exact posterior, {\color{black} and therefore includes uncertainty quantification, access to credible intervals, and Bayesian hypothesis testing}.

\paragraph{Neural-network based posterior estimation}

Recent research indicates that deep neural networks for posterior estimation achieve superior results with fewer simulations than sample-based ABC methods. Two main classes that leverage deep neural network capabilities have been proposed for Bayesian inference.

The first type is based on sequential methods that concentrate on inferring the posterior for individual observations, aiming to optimize simulation efficiency for each data point.
The idea is to parametrically approximate posterior distributions over multiple rounds of adaptively chosen simulations. In each round, a simulator is run using parameters sampled from the current approximate posterior. 
Since drawing simulation parameters from the prior can be wasteful, adaptively chosen proposals can be corrected  
numerically and post-hoc \citep{papamakarios2016fast}  or using importance weights that increase variance during learning \citep{lueckmann2017flexible}.
To overcome optimization problems from many of these sequential methods, \citet{deistler2022truncated} performed sequential inference with truncated proposals.

The second type of approach, known as amortized methods \citep{zammitmangion2024neuralmethodsamortisedinference}, aim to compute posteriors that can be generalized and applied to multiple observations without the need for retraining for each new observation. Normalizing flows has been used to approximate posterior distributions for Bayesian variational inference in \citet{rezende2015variational}, \citet{papamakarios2017masked}, {\color{black} and \citet{papamakarios2021}} without the need to compute parameters per data point, thereby amortizing their estimation. {\color{black} There are also well-documented software packages on neural posterior estimation, such as BayesFlow \citep{radev_et_al_2022, Radev2023}}. As an alternative to invertible neural network approaches such as normalizing flows, \citet{polson2023generative} used implicit quantile neural networks to calculate functionals of interest. {\color{black} Other methods perform posterior estimation alongside an approximation of an intractable likelihood function, such as in \citet{Wiqvist2021SequentialNP}, \citet{glöckler2022variationalmethodssimulationbasedinference}, and \citet{pmlr-v216-radev23a}}.

{\color{black}In this paper, we demonstrate the power and flexibility of a variational assumption on the posterior for amortized posterior estimation, and so the work in this paper would fall into the class of amortized methods.}


\subsection{Our Contribution}


{\color{black}

In this paper, we explore the use of simulation-based inference for environmental applications. In these problems there are often just a few parameters of interest. Hence, approximation of the full joint posterior may not be required; only the marginal posteriors of the parameters of interest. In this case, variational approximations can serve as an alternative to modern simulation-based inference approaches, such as normalizing flows. In this paper, we demonstrate that a variational approach can accurately approximate the marginal posteriors. While the quality of the posterior approximation depends on the variational assumption, Bayesian model checking approaches, such as probability integral transform plots, can be checked and the variational assumption reconsidered.

While it may be possible to approach environmental applications with modern simulation-based inference, a variational approach offers a number of advantages. First, normalizing flows cannot directly model discrete parameters of interest such as the number of infected individuals in a region. On the other hand, a discrete distribution can be chosen as the variational approximation, where hyperparameters are defined as neural networks with the observed data as inputs. Additionally, we were able to obtain accurate results with just 2-3 hidden layers in our neural networks, resulting in less neural network parameters to optimize than modern normalizing flows. Because of this, these neural networks can be fit with fewer simulations, which is significant if it takes a long time to generate simulations. Lastly, even with a large number of simulated datasets, density approximation is difficult in high dimensions such as spatial processes over many locations. We refer to our approach of targeting marginal posteriors with carefully chosen variational approximations as ``variational neural Bayes", or VaNBayes. We show that VaNBayes works well in a wide variety of environmentally and ecologically-inspired applications, and compare with Bayesflow, a modern normalizing-flow based approach to posterior approximation.

}

This paper is structured as follows. Section \ref{s:method} describes VaNBayes, derives its theoretical support and discusses tuning the approximation. Section \ref{s:sim} applies the method to several models, showcasing its wide applicability. Section \ref{s:data analysis} employs VaNBayes to estimate the spread of the Zika Virus throughout the Brazilian states. Section \ref{s:discussion} concludes. The code is available at: \url{https://github.com/macedaell/VaNBayes}. 

\section{The VaNBayes Approach to Amortized NPE}\label{s:method}

\subsection{Model and Estimation}\label{s:framework}

Let $\bY = (Y_1,...,Y_n)^\top$ be the response vector with likelihood function $f(\bY | \btheta)$ depending on the set of model parameters $\btheta = (\theta_1,...,\theta_P)^\top$.  We posit the Bayesian model
\begin{eqnarray}\label{e:model}
   \bY | \btheta &\sim& f(\bY|\btheta)\\
   \btheta &\sim& \pi(\btheta)   
\end{eqnarray}
where $\pi$ is a prior distribution on $\btheta$.  We assume that $f$ and/or $\pi$ are intractable density/mass functions, but that both are straightforward to sample from. We assume that we are only interested in inference on the $Q$-dimensional summary $\bgamma = (\gamma_1,...,\gamma_Q) =  G(\btheta)$ of the parameters $\btheta$. For example, if only the marginal distribution of $\theta_1$ is of interest we would set $Q=1$ and $\gamma = \theta_1$; if the interest is in the $Q$ marginal distributions of $\bgamma$ then the methods below can be applied separately for each $\gamma_j$. The same approach can be used to approximate a posterior predictive distribution by defining $\bgamma$ to be the response for a new observation (see Section \ref{ss:sparse_linear_reg} below). 

Similar to variational Bayesian methods, we assume that the posterior distribution of the parameter of interest follows a parametric distribution.  Define
\begin{equation}\label{e:het_normal}
    \bgamma|\bY \sim p \left\{\bgamma| a(\bY;\bW)\right\},
\end{equation}
where $p$ is known or selected by the user, $a(\bY;\bW) = \{a_1(\bY;\bW_1),...,a_J(\bY;\bW_J)\}$, $\bW = \{\bW_1,...,\bW_J\}$ and each of the $a_j$ is a {\color{black} machine learning model} with {\color{black} parameters} $\bW_j$ that must be learned to mimic the map of the data to the marginal posterior distribution of $\bgamma$. {\color{black}Note that the parametric family of this distribution, $p$, is not assumed to be Gaussian and instead left to the user.}

For example, if $Q=1$ and $\gamma$ has support on the real line then, under the conditions of the Bernstein-von Mises theorem, a reasonable choice is the heteroskedastic Gaussian model
with mean and variance that depend on $\bY$. {\color{black}Alternatively, if we are not confident the Bernstein-von Mises theorem holds and we know that $\gamma$ must be strictly positive, then a reasonable choice could be the gamma distribution, or any other family with a strictly positive support.}
Ultimately, the correct choice of parametric distribution depends on the posterior of interest, $\bgamma$, and its relationship to the data, $\bY$. For example, in the Bayesian {\color{black}sparse} linear regression example in Section 3.2, if $Q=1$ and $\gamma\in\{0,1\}$ is an indicator variable determining the inclusion of a particular covariate in the model, we model $\gamma | \bY \sim Bernoulli(a(\bY))$, where the output of the machine learning model, $a(\bY) \in (0,1)$.

{\color{black} The form of the machine learning model is up to the user, but our numerical illustrations in Section \ref{s:sim} use a neural network with weights $\bW$ since they are flexible, can efficiently handle a large number of input variables (here, $\bY$), and they make predictions efficiently. In the cases we have explored, only shallow neural networks are required since we limit the focus to low-dimensional parameters and assume a parametric model, so it may be possible to perform this estimation without deep learning.}

To simplify the {\color{black} training}, let $\bZ = S(\bY) = \{S_1(\bY),\ldots,S_m(\bY)\}$ be a summary statistic that captures the features in $\bY$ that are important for estimating $\bgamma$. With this pre-processing, we can write $a_j(\bY;\bW_j) = a_j(\bZ;\bW_j)$ and $a(\bY;\bW) = a(\bZ;\bW)$.  Ideally, $\bZ$ is a low-dimensional sufficient statistic, but in other cases these could be user-selected approximations. It is also possible to retain the entire dataset, $\bZ=\bY$, if no suitable dimension reduction can be postulated.

{\color{black}Like other amortized neural posterior estimators,} the functions $a(\bZ;\bW)$ are learned based on simulations from the model.  The proposed algorithm begins by sampling $N$ independent draws of $\btheta$ from a training distribution $\Pi$ and then simulating a dataset from each parameter set, 
\begin{equation}\label{e:draws}
 \btheta_i\iid \Pi(\btheta) \mbox{\ \ and \ \ } \bY_i|\btheta_i\indep f(\bY|\btheta_i)\mbox{\ \ for\ \ } i \in\{1,...,N\}.
\end{equation}
The training distribution should have the same support as the prior distribution, but as discussed in Section \ref{ss:proposal} below there is freedom in choosing its exact form. If $\pi$ is intractable, then $\Pi (\cdot) = \pi (\cdot)$ could be selected so as to bypass importance-weighting in the estimation, described next. 

For each simulated dataset, we reduce $\btheta_i$ to $\bgamma_i=G(\btheta_i)$ and $\bY_i$ to $\bZ_i=S(\bY_i)$, and compute the weights $w_i = \pi(\btheta_i)/\Pi(\btheta_i)$. {\color{black} Note that when the proposal distribution, $\Pi(\cdot)$, has thinner tails than the prior, $\pi(\cdot)$, then these weights tend to have high variance. In this paper we take the proposal distribution to be the prior distribution so this is not an issue, but approaches have been proposed to stabilize the weights \citep{Dax_Neural_IS_for_Rapid, Vehtari_pareto_smoothed_IS}.} We maximize the weighted posterior distribution of $\bgamma$ with outcomes $\bgamma_i$, {\color{black} inputs} $\bZ_i$, unknowns $\bW$ and weights $w_i$ to approximate the relationship between the data and the posterior distribution. That is,
\begin{equation}
\label{e:wMLE}
  \widehat{\bW} = \{\widehat{\bW}_1,...,\widehat{\bW}_J \} = \underset{\bW}{\mathrm{argmax}}
\sum_{i=1}^Nw_i\log\left[p\left\{\bgamma_i|a(\bZ_i;\bW)\right\}\right].
\end{equation}
Equation \eqref{e:wMLE} is a type of importance-weighted empirical risk minimization \citep{shimodaira2000improving}. It reweights the log-likelihood to recalibrate the training distribution $\Pi$ to the prior distribution $\pi$. Once this model is trained on simulated data, posterior inference is straightforward.  We simply evaluate the $a_j$ at the observed data to give the approximate posterior, 
\begin{equation}\label{e:fitted_pdf}
\bgamma|\bY \sim p \{\gamma| a(\bZ;\widehat{\bW})\},
\end{equation}
where $\bZ = S(\bY)$ are the observed summary statistics. The computational speed-up using VaNBayes is realized in this posterior inference step, where performing Bayesian inference takes the same computational cost as new predictions. {\color{black}For clarity, the algorithm for training VaNBayes is given in Algorithm \ref{alg:VaNBayes}.}

\begin{algorithm}
\caption{VaNBayes}\label{alg:VaNBayes}
\begin{algorithmic}[1]
\Require Observed data $\bY_0$, likelihood function $f(\bY|\btheta)$, training distribution $\Pi(\btheta)$ and  prior distribution $\pi(\btheta)$. Let $\bgamma = G(\btheta)$ be the parameter of interest and $\bZ = S(\bY)$ be summary statistics. Propose a variational posterior for $\bgamma$ as $p\{\bgamma | a(\bZ;\bW)\}$ for {\color{black} machine learning model} $a(\bZ; \bW)$.
\For{$i=1,\ldots,N$}
\State Generate $\btheta_i \sim \Pi(\btheta)$ and then $\bY_i \sim f(\bY | \btheta_i)$
\State Compute $\bgamma_i = G(\btheta_i)$ and $\bZ_i = S(\bY_i)$
\EndFor
\State Select $\widehat{\bW} = \underset{\bW}{\mathrm{argmax}}\sum_{i=1}^N\frac{\pi(\btheta_i)}{\Pi(\btheta_i)}\log[p\{\bgamma_i | a(\bZ_i; \bW)\}]$
\State Return $p\{\bgamma | a(\bZ_0; \widehat{\bW})\}$ to approximate the true posterior of $\bgamma$.
\end{algorithmic}
\end{algorithm}

\subsection{Choosing the proposal distribution}\label{ss:proposal}

The prior distribution $\pi(\btheta)$ is a natural choice for the training distribution $\Pi(\btheta)$ and used throughout this paper. In this case, $(\btheta_i, \bY_i)$ are draws from the joint distribution and $\btheta_i | \bY_i$ are draws from the target posterior distribution, so it is clear that training the neural network with weights $w_i = 1$ will approximate the target posterior. If the prior is diffuse, however, then few of the simulated $\bY_i$ will resemble the observed data and so training near the observed data will be inefficient.  In this case, selecting a training distribution more similar to the posterior may improve efficiency. One possibility is to sequentially refine $\Pi$ based on preliminary model fits so that $\Pi$ converges to the posterior \citep{lenzi2023towards}. 

Fortunately, the following theorem suggests that the procedure is insensitive to the training distribution if the number of Monte Carlo replicates $N$ is large. 

\begin{theorem}\label{thm:1} Given that $\Pi(\btheta)$ is a valid distribution of $\btheta$ with the same support as the prior for our model, $\pi(\btheta)$, the weights $\widehat{\bW}$ chosen by VaNBayes are asymptotically (in $N$) invariant to the training distribution used, $\Pi(\btheta)$. For large Monte Carlo replicates $N$, the VaNBayes weights $\widehat{\bW}$ are chosen as if the prior distribution $\pi(\btheta)$ was used as the training distribution.
\end{theorem}
The proof of Theorem \ref{thm:1} is given in the Appendix. 

\subsection{Theoretical Guarantees/Choosing the variational posterior distribution}

Theorem \ref{thm:2} below characterizes the {\color{black} quality} of the VaNBayes approximation {\color{black} to} the true posterior {\color{black} in terms of Kullback-Leibler divergence}. 

\begin{theorem} \label{thm:2} Let $m_0(\bZ)$ be the marginal distribution of the summary statistic data and $p\{\gamma | a(\bZ; \bW)\}$ be the {\color{black} assumed} posterior distribution {\color{black} with parameters modeled by a machine learning model}. The weights chosen by VaNBayes, $\widehat{\bW}$, converge in probability to (possibly non-unique) weights that  minimize the Kullback-Leibler divergence between the (true) joint posterior and $p\{\bgamma | a(\bZ; \bW)\}m_0(\bZ)$ as the number of training samples diverges, {\color{black} $N\to\infty$}.
\end{theorem}
\begin{proof}
\bigskip
\noindent
Denote the prior distribution for $\bgamma$ induced by the prior for $\btheta$ as $\pi_0(\bgamma)$ and the true joint, marginal, and posterior distributions as $p_0(\bZ, \bgamma) = f(\bZ|\bgamma)\pi_0(\bgamma)$, $m_0(\bZ) = \int p_0(\bZ, \bgamma)d\bgamma$, and $p_0(\bgamma|\bZ) = p_0(\bZ, \bgamma)/m_0(\bZ)$, respectively. Let $D_{KL}$ denote the Kullback-Leibler divergence. Then, with all expectations taken with respect to $p_0(\bZ, \bgamma)$, 
\begin{align*}
\widehat{\bW} & = \underset{\bW}{\mathrm{argmax}} \sum_{i=1}^N \frac{\pi(\btheta_i)}{\Pi(\btheta_i)}\log[p\{\bgamma_i | a(\bZ_i; \bW)\}] \\
& \overset{p}{\to} \underset{\bW}{\mathrm{argmax}} \ \E\log[p\{\bgamma | a(\bZ; \bW)\}] \\
& = \underset{\bW}{\mathrm{argmax}} \ \E (\log[p\{\bgamma | a(\bZ; \bW)\}] +\log[m_0(\bZ)]- \log[p_0(\bgamma, \bZ)])\\
& = \underset{\bW}{\mathrm{argmin}} \ \E\left(\log\left[\frac{p_0(\bgamma, \bZ)}{p\{\bgamma | a(\bZ; \bW)\}m_0(\bZ)}\right]\right)\\
& = \underset{\bW}{\mathrm{argmin}} \ D_{KL}[p_0(\bgamma | \bZ)m_0(\bZ) \ || \ p\{\bgamma | a(\bZ; \bW)\}m_0(\bZ)]. \qquad \qedhere
\end{align*}

\end{proof}

This gives some insight as to how the weights are chosen. In Variational Bayes, the variational posterior's {\color{black} parameters} are chosen such that the variational posterior resembles the true posterior. In VaNBayes, the quantity $m_0(\bZ)p\{\bgamma| a(\bZ; \bW)\}$ can be interpreted as the joint posterior parametrized with {\color{black} a machine learning model}. These weights are chosen to minimize the ``distance'' (as measured by the Kullback-Leibler divergence) of this neural network posterior parametrization to the true joint posterior. This observation gives some intuition into the amortization that occurs with VaNBayes --- since we optimize over all potential data instead of our observed data, we can fit as many datasets as desired after the {\color{black} machine learning} weights are determined.

{\color{black}If the true posterior follows the assumed parametric model, then Theorem 2 confirms that VaNBayes can provide a close approximation to the true posterior.} Let $\bnu(\bZ)$ denote the functional relationship between the parameters of the posterior distribution and the data used to perform the inference, $\bZ$. That is, $\bgamma | \bZ \sim g\{\bnu(\bZ)\}$ for some distribution $g$. Informally, we could rewrite the posterior as $p\{\bgamma | \bnu(\bZ)\}$. {\color{black} If we use neural networks as the machine learning model,} the universal approximation theorem states that for smooth $\bnu(\bZ)$, there exists some neural network with weights $\bW^*$ such that $\bnu (\bZ) = a(\bZ; \bW^*)$. When this {\color{black} holds, and the correct parametric family is assumed for the posterior}, we have $p\{\bgamma | \bnu(\bZ)\} = p\{\bgamma | a(\bZ; \bW^*)\}$, and since the Kullback-Leibler divergence is nonnegative, the weights $\widehat{\bW}$ are asymptotically chosen such that
\begin{align*}
D_{KL}[p_0(\bZ, \bgamma) \ || \ m_0(\bZ)p\{\bgamma | a(\bZ, \widehat{\bW})\}] & = D_{KL}[p_0\{\bgamma|\bnu(\bZ)\}m_0(\bZ) \ || \ p\{\bgamma | a(\bZ; \widehat{\bW})\}m_0(\bZ)] \\
& = D_{KL}[p_0\{\bgamma|a(\bZ; \bW^*)\}m_0(\bZ) \ || \ p\{\bgamma | a(\bZ; \widehat{\bW})\}m_0(\bZ)] \\
& = 0.
\end{align*}
Hence, $m_0(\bZ)p\{\bgamma | \bZ, a(\bZ; \widehat{\bW})\}$ resembles the joint posterior, and so $p\{\bgamma | a(\bZ; \widehat{\bW})\}$ resembles the posterior for some realized dataset $\bZ$. Note that in this case $\widehat{\bW}$ need not be equal to $\bW^*$, {\color{black} since we are only interested in the quality of predictions of the machine learning model, not the uniqueness of its weights.}

The assumption that $\bnu(\bZ)$ is smooth enough is technically not verifiable, since our working model is assumed to be intractable. However, we may check this assumption empirically --- if $\widehat{\bW}$ is obtained with a large amount of data and the values of $a(\bZ; \widehat{\bW})$ seem to only slightly change with slightly different samples $\bZ$, then this assumption could be reasonable.

\subsection{Evaluating the fit of the posterior approximation}\label{ss:fit}

Selecting the parametric family and neural network architecture for the approximate posterior are critical steps in the proposed method.  Fortunately, unlike a typical Bayesian analysis of a single dataset, we have access to virtually unlimited validation data from additional simulations to compare the fit of different models and evaluate the fit of the final selection.  Let $(\bgamma_v,\bZ_v)$ for $v\in\{1,...,V\}$ be a set of validation data generated separately from the training data but following the same distribution.  We recommend using the log score \citep[e.g.,][]{gneiting2007strictly}, $\mbox{LS} = \sum_{v=1}^V \log p \{ \bgamma_v|a(\bZ_v;\widehat{\bW})\}$,
for model comparison because of its similarity to the cost function. In the examples of Section \ref{s:sim}, we compute the log score for several models and select the one with the largest log score.  

To confirm that the selected model fits the data well, we recommend the probability integral transform (PIT) plot \citep[e.g.,][]{gneiting2007strictly}.  Letting $F$ be the distribution function corresponding to the fitted model in \eqref{e:fitted_pdf}, the PIT statistic for validation observation $v \in \{1, \ldots, V\}$ is $\mbox{PIT}_v = F\{\bgamma_v|a(\bZ_v;\widehat{\bW})\}$.
Assuming the model fits well, the PIT statistics should follow a Uniform(0,1) distribution. We evaluate the fit using a QQ-plot of the empirical distribution of the $\mbox{PIT}_v$ versus the uniform distribution; deviations from the diagonal $x=y$ line suggest a lack of fit. {\color{black}Other options include simulation-based calibration \citep{talts_validating_2020}, which has previously been used in simulation-based inference settings, such as in Bayesflow \citep{radev_jana_2023}. Other newly derived simulation-based inference measures of fit could also be used \citep{anau_montel_tests_2025, schmitt_detecting_2024}.}

If there is evidence of a lack of fit, we then revisit the choice of parametric model and network architecture, and investigate convergence of the optimization algorithm. {\color{black} Often, good approximations can be found after applying suitable transformations, e.g., using a log transformation so that $\gamma$'s distribution is unbounded. As with any model-building exercise, it is incumbent on the user to propose and validate parametric choices, but we find that the heteroskedastic normal model provides a reasonable approximation in most of the applications we have considered.}

\section{Numerical Illustrations}\label{s:sim}

We begin with an evaluation of the proposed method using multiple linear regression. Although simulation-based inference methods are not required in these cases we explore them to compare VaNBayes to Bayesflow \citep{radev_et_al_2022, Radev2023} in Section \ref{ss:MLR_comparison} and MCMC in Section \ref{ss:sparse_linear_reg}.

{\color{black}Bayesflow \citep{radev_et_al_2022} is a modern neural posterior estimator that is constructed using two separate neural networks; a summary network and an inference network. The summary network is used to find informative summary statistics of the input data. This can be excluded from the Bayesflow architecture if we opt to use hand-crafted summary statistics instead. We will do this in many of these examples to best compare VaNBayes's variational assumption against the normalizing flow used in Bayesflow's inferential network. The inferential network is made from a sequence of Affine Coupling Blocks (ACB's), which are nonlinear invertible transformations relating the posterior distribution to the latent multivariate normal distribution. This normalizing flow framework converges to the exact posterior if the training sample is large and the normalizing flow is complex enough.}


We then consider several more complicated models in the remainder of the section. We use the simulation study to determine whether the proposed method gives appropriate frequentist properties including small bias of the posterior mean estimates and nominal coverage of posterior credible sets. We also study sensitivity to the structure of the posterior approximation and show how to compare and evaluate fitted models.

\subsection{Multiple Linear Regression}\label{ss:MLR_comparison}

In this simulation study we compare VaNBayes to Bayesflow \citep{radev_et_al_2022, Radev2023} in linear regression with varying dimensions. Although this model is relatively straightforward, the purpose is to demonstrate that other simulation-based applications can be inefficient if they compute the entire posterior when only few parameters are of interest. A key difference between VaNBayes and other simulation-based inference methods is the ability to explicitly target low-dimensional {\color{black}marginals} of the posterior. In this example, we model the marginal posterior distributions, decreasing the computational cost needed to estimate the parameters of interest.

The linear regression data-generating model is 
\begin{equation}\label{e:sparse_lin_reg}
Y_i = \beta_0 + \sum_{j=1}^pX_{ij}\beta_j + \varepsilon_i
\end{equation}
with $\varepsilon_i\iid\mbox{Normal}(0,\sigma^2)$ for $i\in\{1,...,n\}$. We set $n=100$
and generate the $p$ covariates $X_{ij}$ from independent standard normal distributions.
The true values of the parameters are set to $\beta_0 = 1$, $\beta_j = 2[(j \ \mbox{mod} \ 5) - 2]$ and $\sigma^2$ is chosen so the signal-to-noise ratio is 0.8 for each simulation. We use uninformative priors $\beta_0, \beta_j \overset{iid}{\sim} \mbox{N}(0,10)$ and $\sigma^2 \sim \mbox{InvGamma(0.50,0.05)}$. We use the least squares  estimates for the coefficients and the log-transformed estimate of $\sigma$ as summary statistics $\bZ$ for both VaNBayes and Bayesflow. We use these $p+2$ statistics as inputs to two-layered VaNBayes networks for each covariate and a two-layered Bayesflow inferential network. Each network is trained with $N=8000$ simulations and validated with $2000$ simulations. We consider $p \in \{ 5, 10, \ldots, 40 \}$.

\begin{figure}
\centering 
\includegraphics[width=0.7\textwidth,page=3]{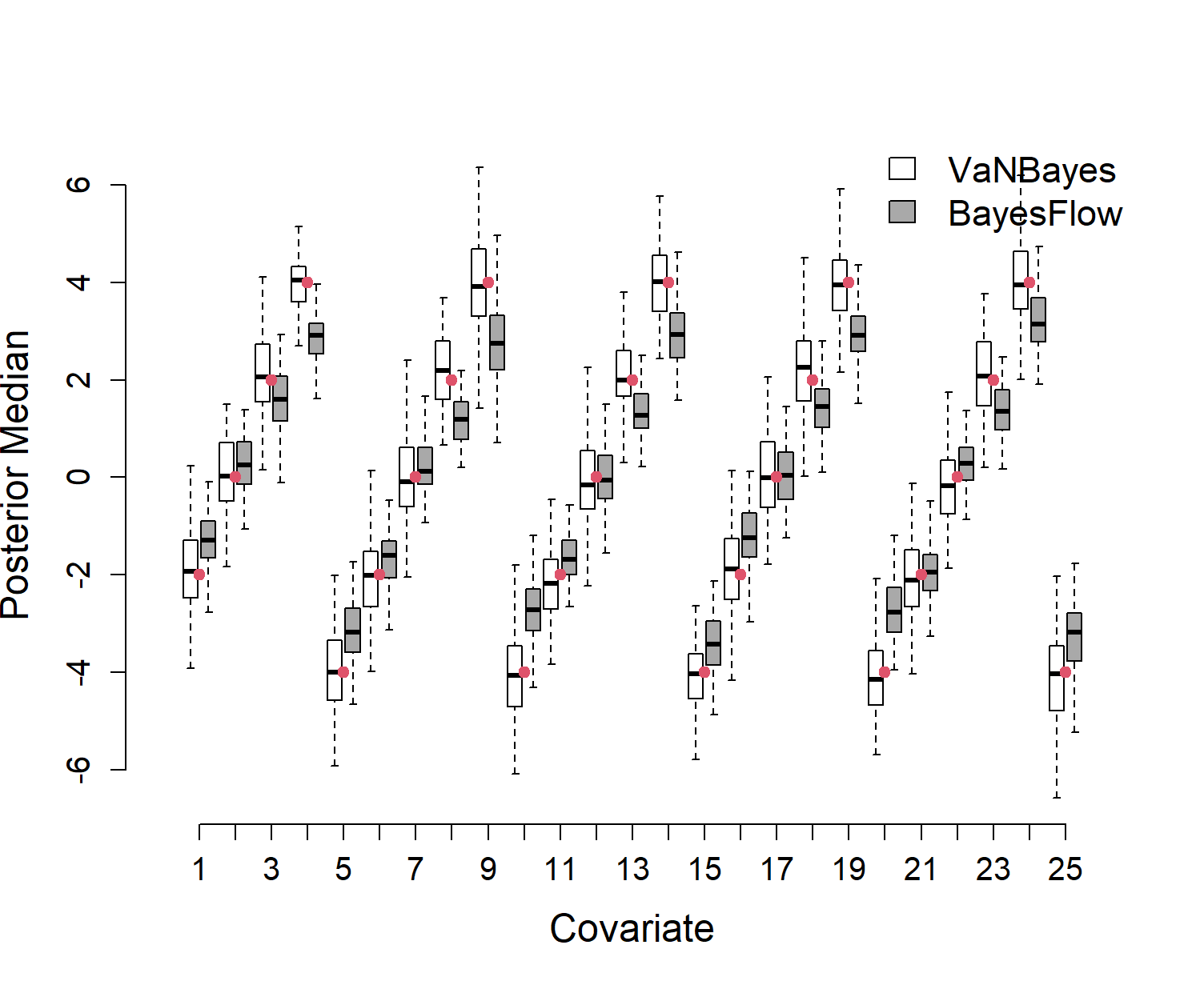}
\caption{The posterior medians of Bayesflow and VaNBayes approximate marginal posteriors of each of the covariates with the true value (red dot) in the $p=25$ case.}
\label{f:posterior_medians_compare}
\end{figure}

\begin{figure}
\centering 
\includegraphics[width=0.5\textwidth,page=1]{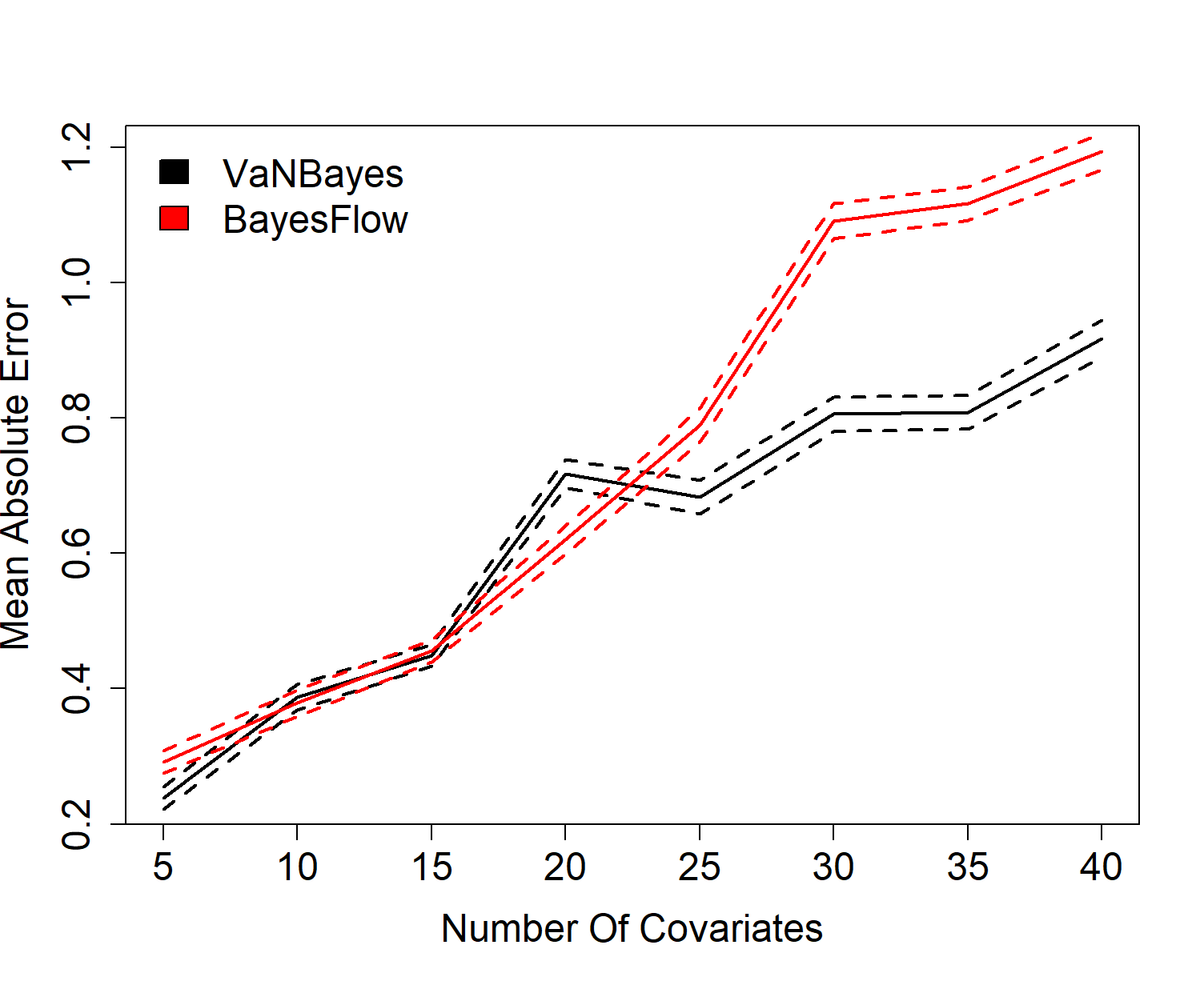}
\caption{The mean absolute error averaged across the covariates for VaNBayes and Bayesflow as a function of the number of covariates, $p$. The dashed lines are the 95\% confidence intervals.}
\label{f:MAE_compare}
\end{figure}

Figures \ref{f:posterior_medians_compare} and \ref{f:MAE_compare} compare the results of VaNBayes and Bayesflow. In Figure \ref{f:posterior_medians_compare}, Bayesflow posterior medians tend to have less variance, but sometimes biased from the true values. On the other hand, VaNBayes posterior medians seem to be unbiased, but with larger variance. In Figure \ref{f:MAE_compare}, as $p$ increases the average mean absolute error (MAE) of both methods increase, but Bayesflow's MAE increases more dramatically. This simple example shows the benefit of VaNBayes avoiding approximation of the entire posterior distribution. 

\subsection{Sparse linear regression}\label{ss:sparse_linear_reg}

The sparse linear regression data-generating model is also (\ref{e:sparse_lin_reg}).  The $p$ covariates are generated as Gaussian with mean zero, variance one and $\mbox{Cor}(X_{ij},X_{ik}) = \rho^{|j-k|}$.  The true values of the parameters are set to $\beta_0=0$, $\beta_1=\beta_2=\beta_6=0.5$,  $\beta_j=0$ all other $j$, $\rho=0.5$ and $\sigma=1$.  We simulate 100 datasets from this model with $p=10$ and $p=20$, each with $n=50$. For each simulated dataset, we also {\color
{black} approximate the posterior predictive distribution of} $10$ test set observations, $Y_{n+1},\ldots,Y_{n+10}$ from the same model to evaluate out-of-sample prediction.  

The prior distribution is based on \cite{george1993variable}.  The sparsity prior for the regression coefficients is the two-component mixture distribution with $\beta_j = 0$ with probability $1-\pi$ and $\beta_j\sim\mbox{Normal}(0,\tau^2)$ with probability $\pi$, independently for $j\in\{1,...,p\}$.  This prior assigns probability $1-\pi$ to the event that covariate $j$ is null and removed from the model.  The remaining priors are $\beta_0\sim\mbox{Normal}(0,v^2)$, $\sigma^2\sim\mbox{InvGamma}(a,b)$ and $\pi\sim\mbox{Beta}(c,d)$.  For this small example, we use weakly informative priors by setting $v=\tau=1$, $a=0.5$, $b=0.05$ and $c=d=2$, which gives prior 95\% interval $(0.14,10.09)$ for $\sigma$ and $(0.09, 0.91)$ for $\pi$. Our objective is to estimate the error standard deviation, $\sigma$, and the posterior inclusion probabilities given $\bY=(Y_1,...,Y_n)$, $\mbox{PIP}_j = \mbox{Prob}(\beta_j\ne 0|\bY)$ for $j \in \{1,...,p\}$.

Each simulated dataset is analyzed using MCMC and the proposed VaNBayes methods. We do not include Bayesflow in this example because, to our knowledge, it does not apply to discrete posterior distributions. For MCMC, we use Gibbs sampling with the true parameter values as initial values and $40{,}000$ iterations after a burn-in of $10{,}000$ iterations. For the proposed VaNBayes methods, $N=100{,}000$ datasets were simulated from the model in \eqref{e:sparse_lin_reg} with parameters simulated from the prior.  We apply the proposed methods to approximate the posterior distribution of the $p$ marginal posterior inclusion probabilities, $\mbox{PIP}_j$, the marginal posterior of the error standard deviation, $\sigma$, and the posterior predictive distribution of 10 test set observations simulated following the same distribution as the observed data.  For all parameters and predictions, the $p+3$ summary statistics in $\bZ$ are the least squares estimates of $(\beta_0,...,\beta_p)$, the residual standard deviation and the standard deviation of the least squares estimates.  The summary statistics are all rank transformed to $[-1,1]$.  

For each simulated dataset, we extract $\gamma_j = \mathbbm{1}(\beta_j\ne 0)\in\{0,1\}$ for $j\in\{1,...,p\}$, $\gamma_{p+1} = \sigma$ and {\color{black} ten posterior predictive distributions} $\gamma_{p+1+i} = Y_{n+i}$, $i\in \{1, \ldots, 10\}$.    Separately from each $j\in\{1,...,p\}$, we fit the logistic regression model
$\mbox{logit}\{\mbox{Prob}(\gamma_j=1|\bZ)\} = a_j(\bZ;\bW_j)$, where $a_j(\bZ;\bW_j)$ is a feed-forward neural network (FFNN) with inputs $\bZ$ and two layers comprised of $L_1$ and $L_2$ nodes, respectively.  We use cross-entropy loss and the ADAM optimizer and the {\tt keras} package in {\tt R} with default settings for all tuning parameters (mini-batch size, learning rate, etc.).  Then $\mbox{PIP}_j$ is taken to be the fitted value/probability from the trained neural network with the observed $\bZ$ as input.   For $\gamma_{p+1} = \sigma$, the proposed VaNBayes method assumes the marginal posterior distribution follows a log-normal distribution 
$$\sigma| \bY \sim \mbox{logNormal}\left[A_1(\bZ;\bw_1),\exp\{A_2(\bZ;\bw_2)\}\right].$$
The networks $A_1$ and $A_2$ both have inputs $\bZ$ and two hidden layers with $L_1$ and $L_2$ nodes, but with separate weight parameters, $\bw_1$ and $\bw_2$.  The assumed model for prediction is 
$$Y_{n+i}| \bY \sim \mbox{Normal}\left[B_i(\bZ;\bu_i),\exp\{C_i(\bZ;\bv_i)\}\right]$$
for networks $B_i$ and $C_i$ and weights $\bu_i$ and $\bv_i$.

We use cross-validation over a validation set of size $V=100,000$ to select $L_1$ and $L_2$ (Table \ref{t:pips_cv}).  For validation set observation $v$ and covariate $j$, let $\gamma_{vj}$ be the binary indicator that covariate $j$ is included in the model and $\widehat{\mbox{PIP}}_{vj}$ be the fitted probability from the deep learning.  The metrics are cross entropy loss, classification accuracy and Brier score,
\begin{eqnarray}
    \mbox{CE} &=& -\frac{1}{Vp}\sum_{v=1}^V\sum_{j=1}^p\gamma_{vj}\log(\widehat{\mbox{PIP}}_{vj}) + (1-\gamma_{vj})\log(1-\widehat{\mbox{PIP}}_{vj}),\\
    \mbox{CA} &=& \frac{1}{Vp}\sum_{v=1}^V\sum_{j=1}^p\gamma_{vj}\mathbbm{1}({\widehat{\mbox{PIP}}}_{vj}<0.5) + (1-\gamma_{vj})\mathbbm{1}(\widehat{\mbox{PIP}}_{vj}<0.5),\\
    \mbox{BS} &=& \frac{1}{Vp}\sum_{v=1}^V\sum_{j=1}^p(\gamma_{vj}-\widehat{\mbox{PIP}}_{vj})^2,
\end{eqnarray}
respectively. For $\sigma$ and for prediction, we use the log score as the metric for comparison.  The results for the posterior inclusion probabilities in Table \ref{t:pips_cv} are similar for all network sizes, and so we select the values of $L_1$ and $L_2$ that maximize $\mbox{CA}$. The results are similarly insensitive for $\sigma$ and prediction and are thus not shown. 

\begin{table}
\centering
 \begin{tabular}{ll|ccc|ccc} 
&& \multicolumn{3}{c}{$p=10$} & \multicolumn{3}{c}{$p=20$}\\
 $L_1$ & $L_2$ & CE & CA & BS & CE & CA & BS \\
 \hline
50  & 10 & 0.2939 & 0.8644 & 0.0936 & 0.3058 & 0.8573 & 0.0978\\
50  & 25 & 0.2932 & 0.8646 & 0.0934 & 0.3061 & 0.8570 & 0.0980\\
100 & 10 & 0.2933 & 0.8648 & 0.0934 & 0.3057 & 0.8571 & 0.0978\\
100 & 25 & 0.2934 & 0.8646 & 0.0935 & 0.3057 & 0.8573 & 0.0978\\
200 & 10 & 0.2917 & 0.8652 & 0.0930 & 0.3057 & 0.8574 & 0.0978\\
200 & 25 & 0.2930 & 0.8646 & 0.0934 & 0.3051 & 0.8575 & 0.0976
\end{tabular}
 \caption{Cross-validation error for the sparse linear model with $p$ covariates.  The networks vary by the number of nodes in the two hidden layers ($L_1$ and $L_2$) and are compared using validation set cross-entropy loss (CE), classification accuracy (CA) and Brier score (BS).\label{t:pips_cv}}
\end{table}

\begin{figure}
\centering 
\includegraphics[width=0.48\textwidth]{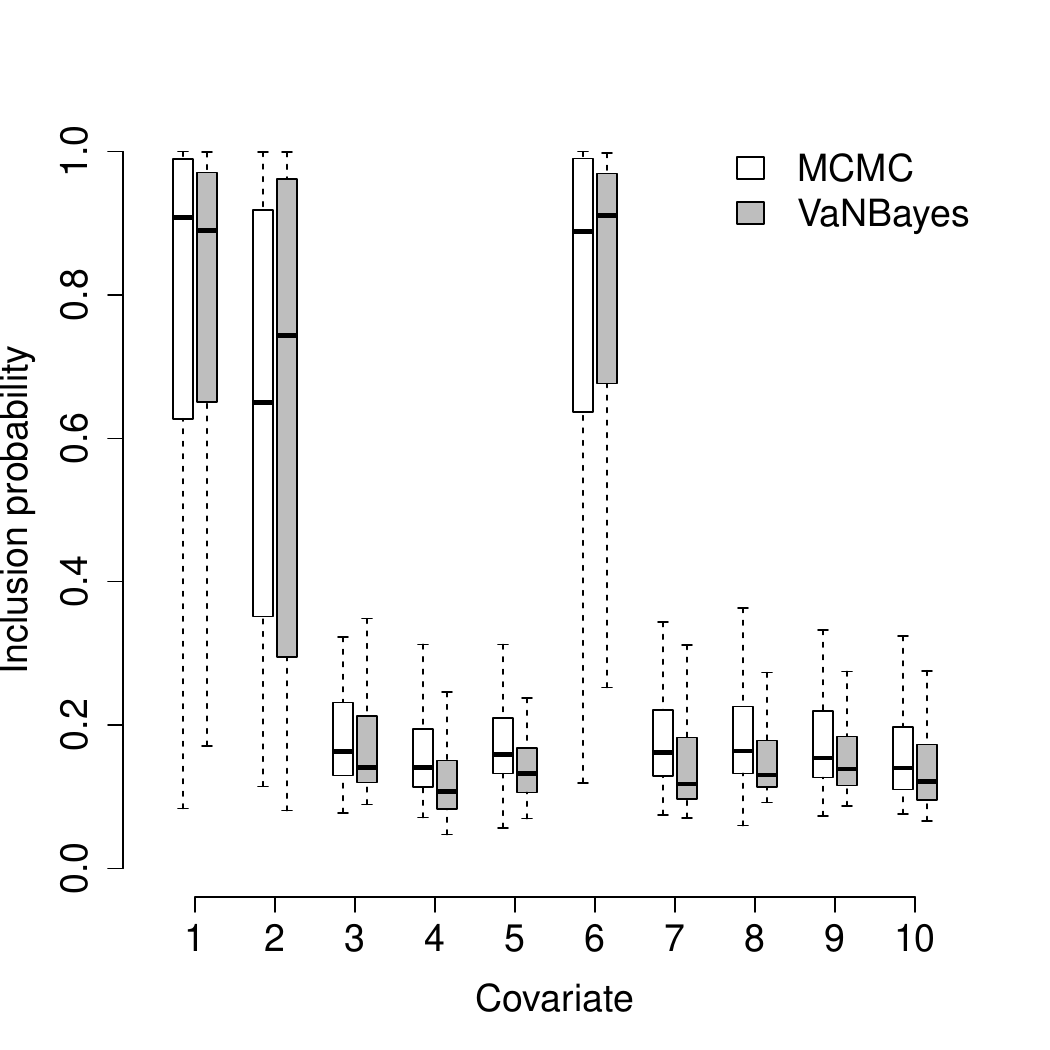}
\includegraphics[width=0.48\textwidth]{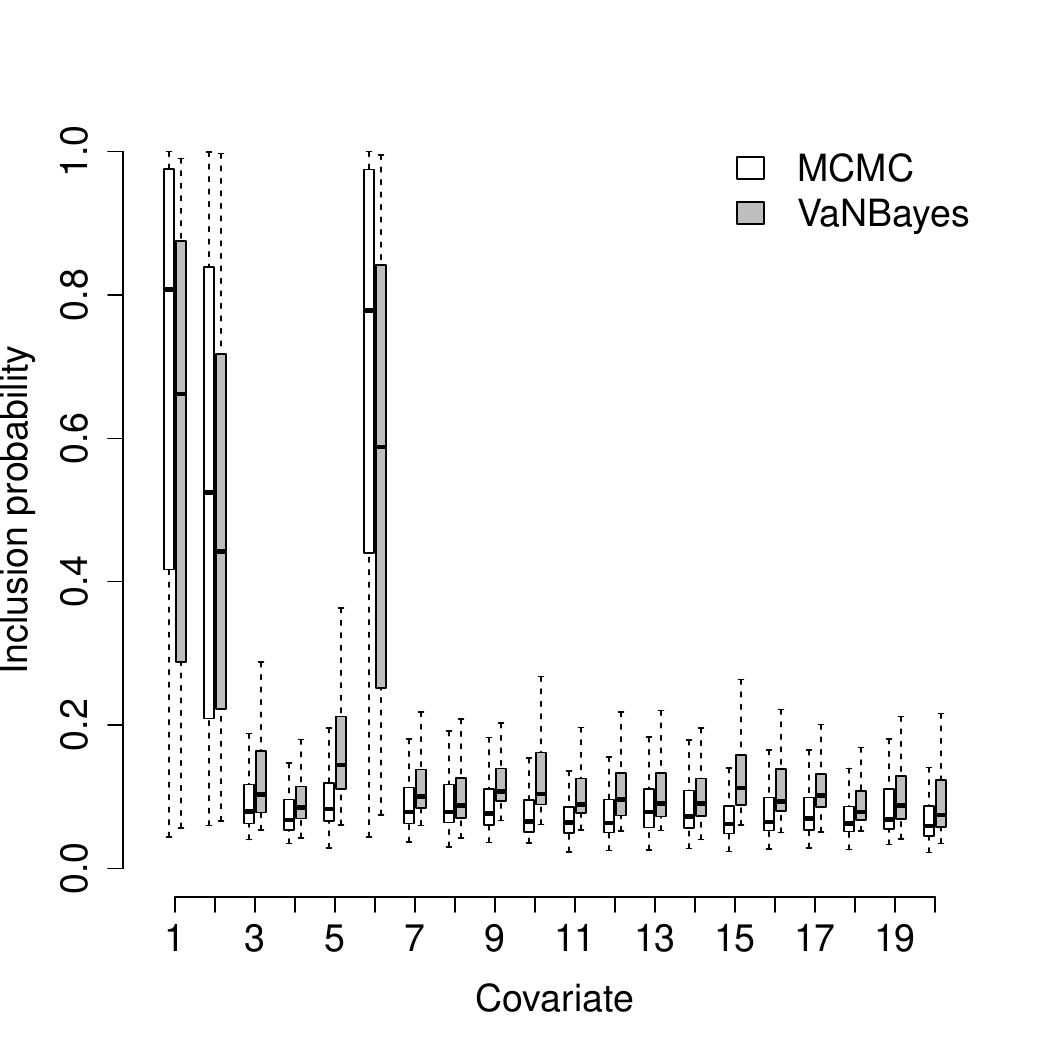}
\caption{Sampling distribution of the posterior inclusion probabilities ($\mbox{PIP}_j$) from MCMC versus the proposed VaNBayes method over 100 simulated datasets from the sparse linear regression model with $p=10$ (left) and $p=20$ (right) and true model that include only variables 1, 2 and 6.\label{f:pips}}
\end{figure}

Figure \ref{f:pips} shows the sampling distribution of $\mbox{PIP}_j$ from both computational approaches.  There is general agreement between the two.  The largest discrepancy is that the VaNBayes approach underestimates $\mbox{PIP}_6$ for $p=20$.  In addition to having similar overall performance, the two methods tend to produce similar estimates on individual datasets.  Pooling $\mbox{PIP}_j$ estimates across covariates and datasets, the correlation between the two estimators is 0.97 for $p=10$ and 0.90 for $p=20$.   

\begin{figure}
\centering 
\includegraphics[width=0.48\textwidth,page=1]{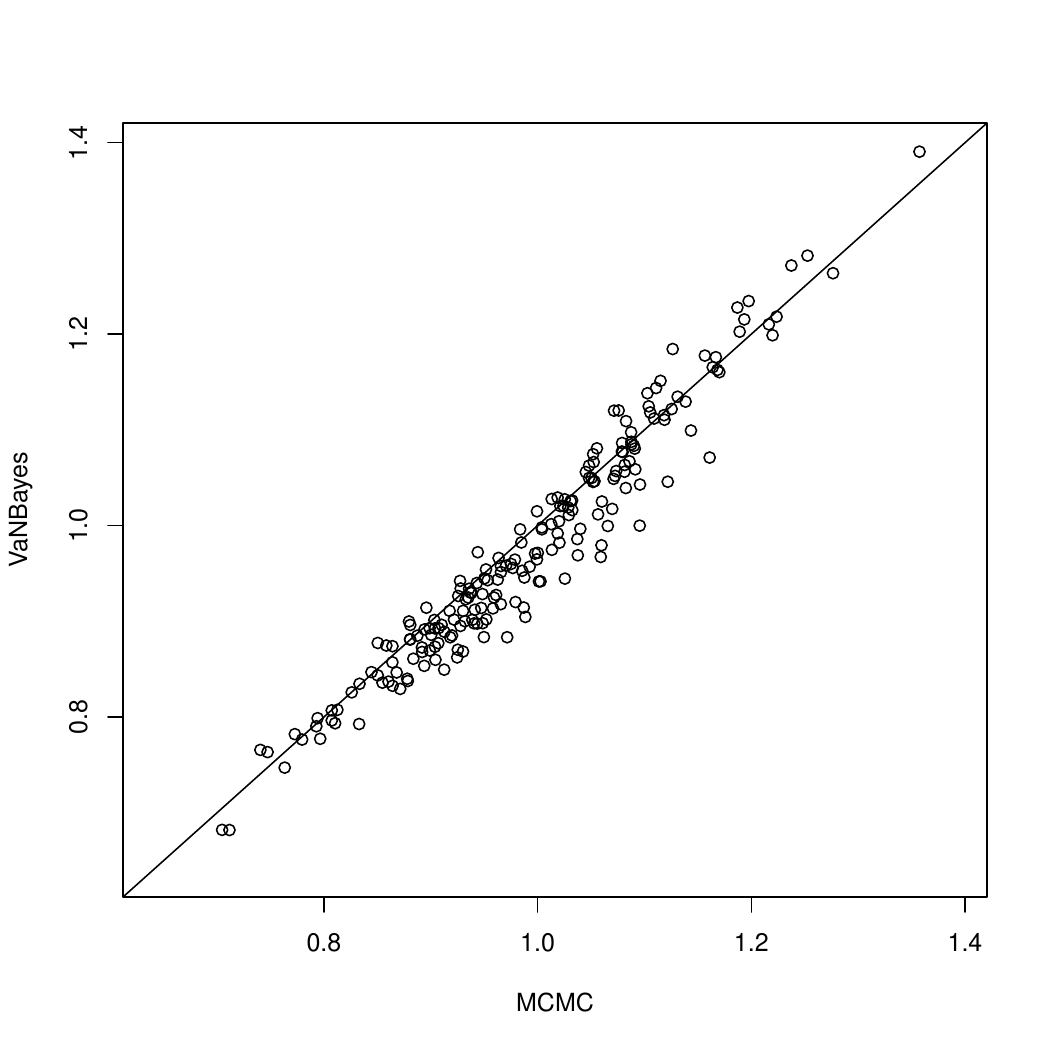}
\includegraphics[width=0.48\textwidth,page=2]{figs/sigma_medians.pdf}
\caption{Posterior median of $\sigma$ for the proposed method and MCMC for $p=10$ (left) and $p=20$ (right).  Each point is one simulated dataset.}
\label{f:mean_sigma}
\end{figure}

Table \ref{t:slm_sigma} gives results for the error standard deviation and prediction.  Methods are compared using the median absolute deviation (MAD) of the posterior median estimator and coverage of 90\% credible sets; for prediction these metrics are averaged over the test set.  MAD and coverage are similar for both methods.  Figure \ref{f:mean_sigma} plots the posterior median estimator from MCMC and the VaNBayes method over the simulated datasets. As with the PIP analysis, the agreement is stronger for $p=10$ than $p=20$, but generally good.  For example, the correlation between posterior medians in Figure \ref{f:mean_sigma} is 0.97 for $p=10$ and 0.89 for $p=20$.  The agreement between VaNBayes and MCMC is similar for prediction.

\begin{table}
\centering
 \begin{tabular}{l|cc|cc|cc|cc} 
 & \multicolumn{4}{c}{Standard deviation} & \multicolumn{4}{c}{Prediction}\\
& \multicolumn{2}{c}{$p=10$} & \multicolumn{2}{c}{$p=20$} & \multicolumn{2}{c}{$p=10$} & \multicolumn{2}{c}{$p=20$}\\
Method & MAD & Cov & MAD & Cov  & MAD & Cov & MAD & Cov\\
 \hline
MCMC  & 0.093 & 0.88 & 0.096 & 0.89 & 0.894 & 0.88 & 0.912 & 0.88\\
VaNBayes   & 0.103 & 0.85 & 0.093 & 0.91 & 0.898 & 0.88 & 0.940 & 0.87\\
\end{tabular}
 \caption{Median absolute deviation (MAD) and coverage of 90\% credible sets (Cov) for the error standard deviation $\sigma$ and test set prediction (averaged over the testing set) in the sparse linear regression model using MCMC and the proposed VaNBayes method.\label{t:slm_sigma}}
\end{table}

\subsection{Autologistic regression model}\label{ss:autologistic}

The autologistic model is an extension of logistic regression to dependent data on a network.  Let $Y_i$ be the binary response and $\bX_i$ the vector covariates (with first element set to one, corresponding to an intercept) for node $i=1,\ldots,n$.  The dependence structure is defined through the edges, with $\mathcal{N}_i$ defined as the collection of indices of the nodes connected to node $i$.  The centered autologistic model of \cite{caragea2014centered} is defined via the full conditional distributions of one node given the others, 
\begin{equation}\label{e:autologistic}
  \mbox{logit}\{\mbox{Prob}(Y_i=1|Y_j, j\ne i)\} = \mbox{logit}(\kappa_i) + \phi\sum_{j\in{\mathcal{N}}_i}(Y_j-\kappa_j),
\end{equation}
where $\mbox{logit}(\kappa_i) = \bX_i^\top \bbeta$ is the usual logistic regression probability with covariate effects $\bbeta=(\beta_1,...,\beta_p)$ and $\phi>0$ determines the strength of dependence.  The priors are standard normal, $\beta_j,\log(\phi)\iid\mbox{Normal}(0,1)$.  While the full conditional distributions have simple forms, the joint likelihood involves a normalizing constant that is the sum of $2^n$ terms and is thus intractable.  However, drawing realizations from the joint distribution is straightforward using, e.g., Gibbs sampling.  

\begin{figure}
\centering 
\includegraphics[width=0.5\textwidth,page=3]{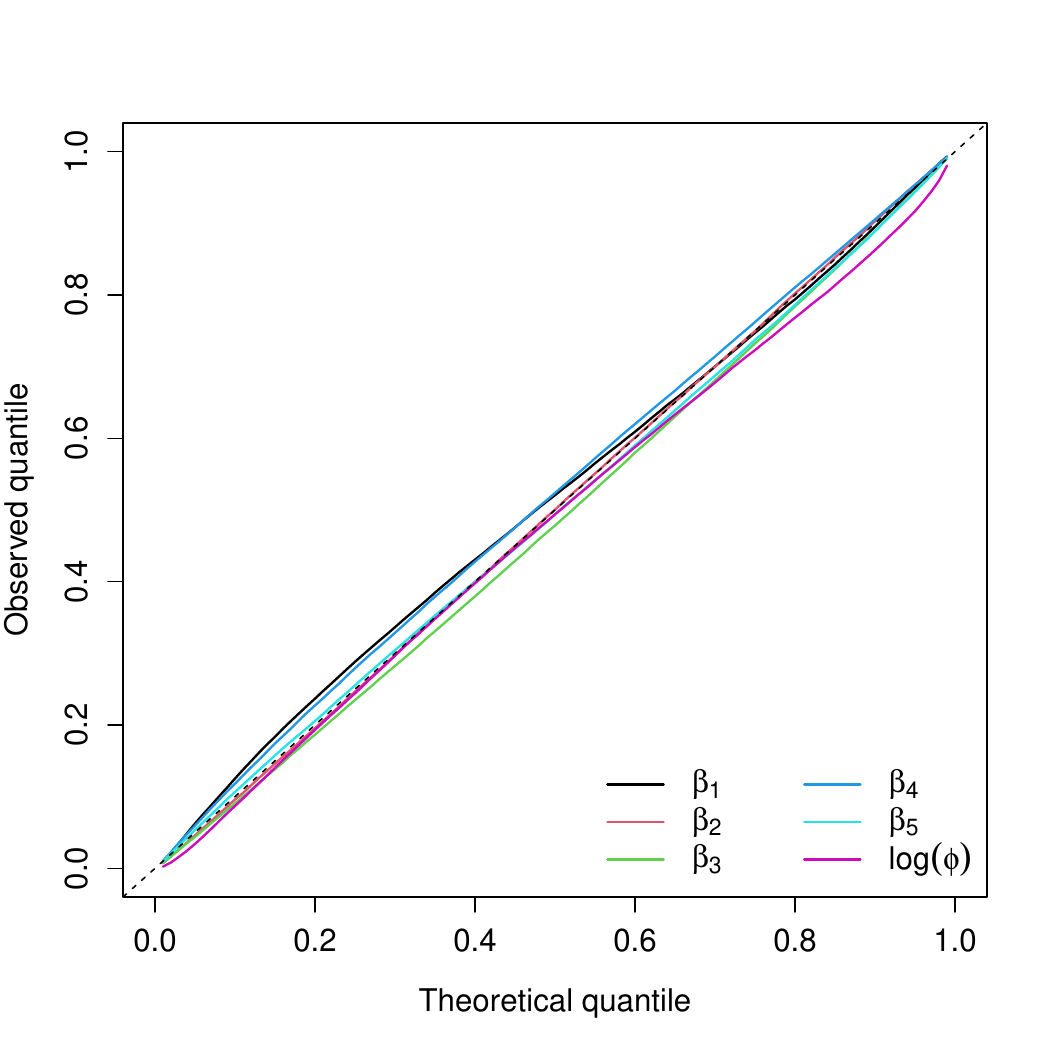}
\caption{QQ-plot of the probability integral transform statistics for the autologistic regression coefficients, $\beta_j$, and log dependence parameter, $\log(\phi)$.}
\label{f:autologistic_pit}
\end{figure}

We simulate $n=400$ observations on a $20\times 20$ grid of regions with rook adjacency and $p=5$ with $X_{ij}\iid\mbox{Normal}(0,1)$.  The $p+1$ parameters of interest are $\bgamma = \btheta=[\beta_1,...,\beta_p,\log(\phi)]$.
We train the model using $N=100,000$ samples with covariates drawn from the standard normal distribution and training distribution for $\btheta$ set to the prior distribution. Define the function $\mbox{expit}(x)=(1+e^{-x})^{-1}$. The $p+3$ summary statistics $\bZ$ are taken to be the non-spatial logistic regression estimate of $\bbeta$, $\widehat{\bbeta}_{GLM}$, and Geary's C statistics \citep{geary1954contiguity} using the residuals $Y_i-\mbox{expit}(\bX_i^\top \widehat{\bbeta}_{GLM})$ and first-, second- and third-order neighbors.  The distribution of each element of $\bgamma$ is modeled using the heterogeneous normal model in \eqref{e:het_normal} and the neural network is trained using the same architecture and tuning parameters as the error standard deviation and predictions in the sparse linear regression case of Section \ref{ss:sparse_linear_reg}.  Figure \ref{f:autologistic_pit} shows that the model fits reasonably well. 

\begin{figure}
\centering 
\includegraphics[width=0.45\textwidth,page=1]{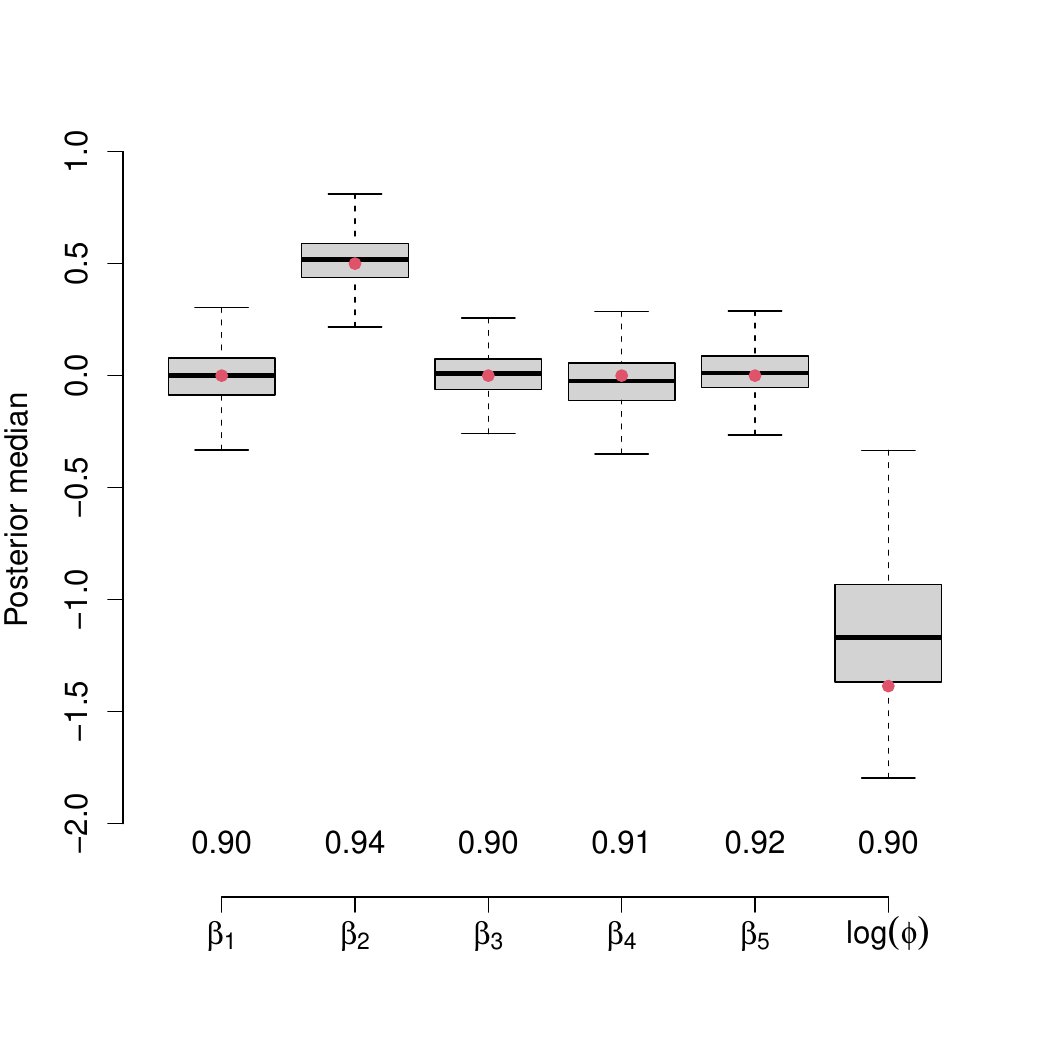}
\includegraphics[width=0.45\textwidth,page=2]{figs/autologistic.pdf}
\caption{Sampling distribution of the posterior median for the autologistic regression coefficients, $\beta_j$, and log dependence parameter, $\log(\phi)$. The panels differ by the true value of $\phi$.  The true values are shown as red dots and the coverage percentages of 90\% posterior intervals are given below the boxplots. }
\label{f:autologistic}
\end{figure}

We simulate 1000 datasets using two sets of true parameters, shown (red dots) in the two panels of Figure \ref{f:autologistic}.  The sampling distribution of the approximate posterior median appears to be unbiased except for the log-dependence parameter in the low-dependence case (Figure \ref{f:autologistic}, left).  However, even in this case, the empirical coverage exceeds the nominal level. 

\subsection{Stochastic differential equations model}\label{s:SIR}
\subsubsection{Nonspatial SIR}

Stochastic differential equations are often used to model the spread of a disease through space and time. The Susceptible-Infected-Removed (SIR) model is a stochastic differential equations model, where the responses are the random variables of the number of infected and recovered individuals across time. Defining an approximate likelihood requires discretizing time, and an accurate approximation results in a likelihood with many terms. It is far easier to simulate from these processes, making simulation-based inference an appealing option.

We employ VaNBayes alongside Bayesflow to demonstrate that VaNBayes achieves similar results to already established methods despite imposing an assumption on the posterior family. {\color{black}Specifically, we will compare VaNBayes and Bayesflow across the number of simulated datasets $N$, which are used as the datasets for VaNBayes and Bayesflow.} This demonstrates that VaNBayes works well as a fast and convenient posterior approximator, {\color{black}even when the marginal posteriors are targeted}. For instance, VaNBayes can get a quick sense of parameter recoverability.

To facilitate the comparison, our implementation of the SIR model in this simulation study is based on the Bayesflow team's implementation and described below. The system of ODE's that govern this process is
\begin{align*}
    \frac{dI}{dt} & = -\lambda \left(\frac{SI}{M}\right) ,\quad
    \frac{dS}{dt}  = \lambda \left(\frac{SI}{M}\right) - \mu I, \quad  
    \frac{dR}{dt}  = \mu I,
\end{align*}
where $M=S+I+R$ is held constant at $M=83\times 10^6$, with initial conditions $R_0 = 0$ and $S_0=N-I_0$. Observations are 14 days of infected counts, $I_0^{(obs)},...,I_{13}^{(obs)}$, where each observed infected count is a Negative Binomial distributed variable according to some dispersion parameter $\psi$, i.e. $I_t^{(obs)}\sim \mbox{NegBinomial}(I_t,\psi)$.

Our simulation study sets the parameters $\lambda = 0.4$, $\mu = .1$, $I_0=20$, and $\psi = 7$. The objective of the simulation study is to approximate the posterior distributions for the parameters of interest, $\lambda$, $I_0$, and $\psi$. These have the respective priors of $\mbox{LogNormal}(\mbox{log}(0.4), .5)$, $\mbox{Gamma}(2,20)$, and $\mbox{Exp}(5)$. These priors were chosen according to the findings in \cite{doi:10.1126/science.abb9789}.

We compare VaNBayes and Bayesflow over 100 datasets generated from the simulation study parameters {\color{black}across $N\in\{500,750,1000,1250,1500\}$. Although these sample sizes are small relative to other simulation-based inference literature, we use these to illustrate the influence of the VaNBayes variational assumption. In this implementation of VaNBayes, we target the marginals of each parameter separately and assume that the marginal posterior of each parameter follows normal distribution.} The parameter $\lambda$ was log-transformed, and $I_0$ and $\psi$ were transformed via 
$
    \theta' = \Phi^{-1}\{ F(\theta) \},
$
where $\theta$ is the original parameter, $F$ is the CDF of the prior distribution, and $\Phi^{-1}$ is the quantile function of the standard normal distribution.

{\color{black} Because VaNBayes targets the marginal posteriors of each parameter separately in this setup, different neural network architectures were used for each parameter. Each neural network used two hidden layers, varying the size of each hidden layer from 10 to 25 and from 5 to 15 respectively. The learning rate for $\lambda$ and $I_0$ were set to $0.001$, and the learning rate for $\psi$ was picked to be $0.0004$. These neural network settings were chosen using the model-fitting tools discussed in Section \ref{ss:fit}. The summary statistics of each dataset were the first seven principal component scores associated with the PCA decomposition, which explain a little over 90\% of the variation on the generated datasets.} To ensure both models used the same input data, Bayesflow was implemented without a summary network, and the inferential network was chosen  to have six layers.

\begin{figure}
\centering 
\includegraphics[width=0.33\textwidth,page=1]{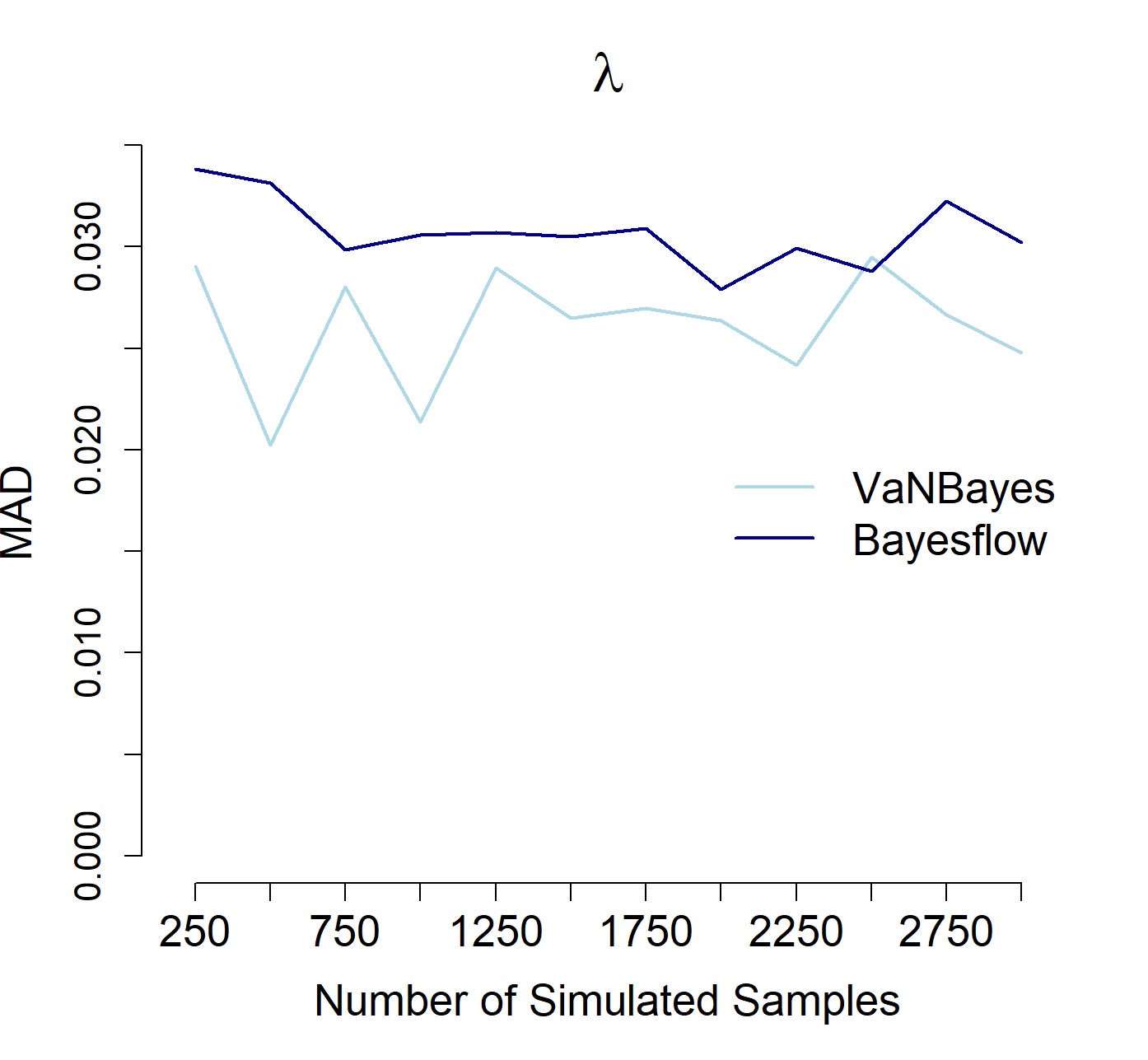}
\includegraphics[width=0.33\textwidth,page=1]{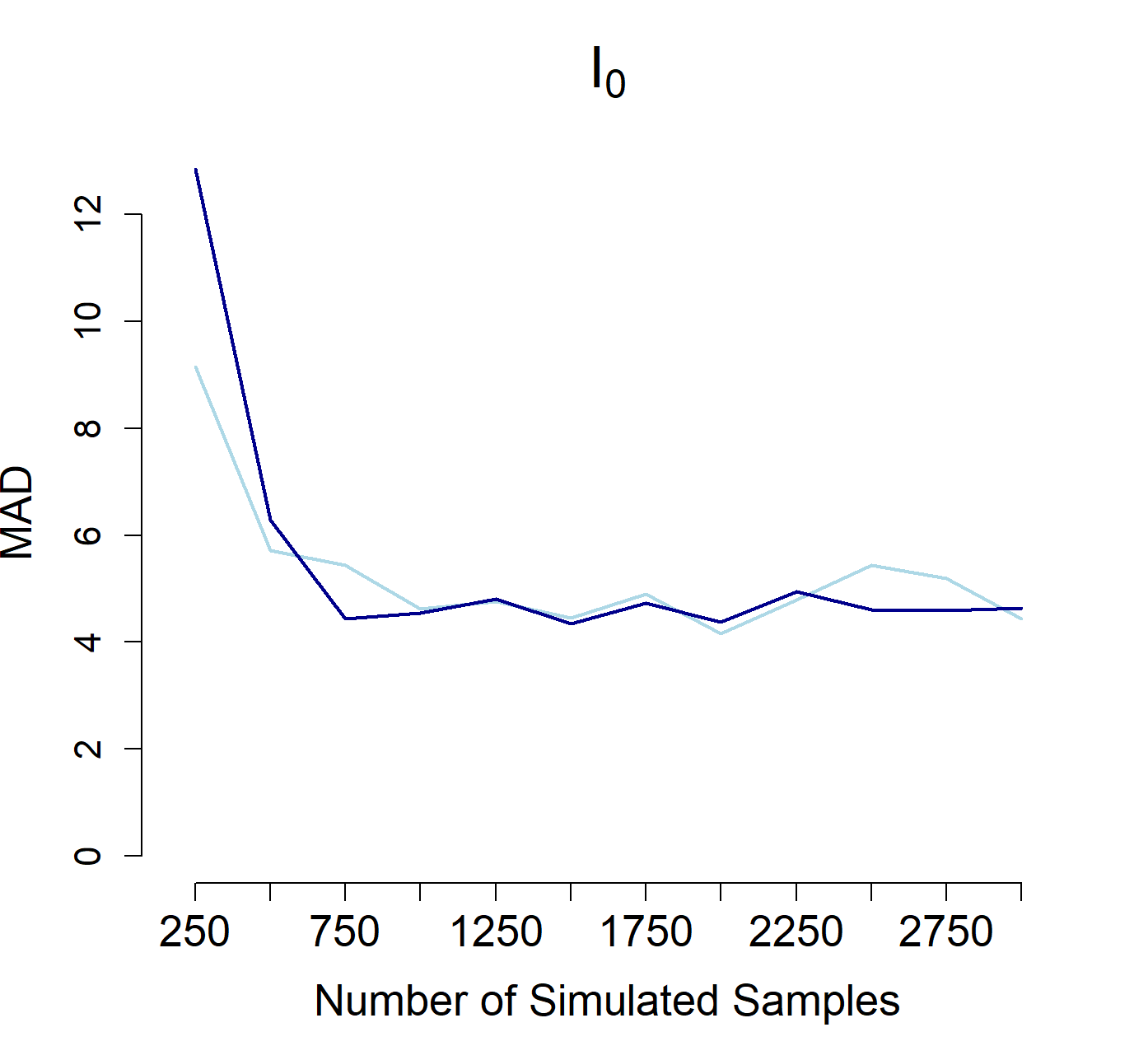}
\includegraphics[width=0.33\textwidth,page=1]{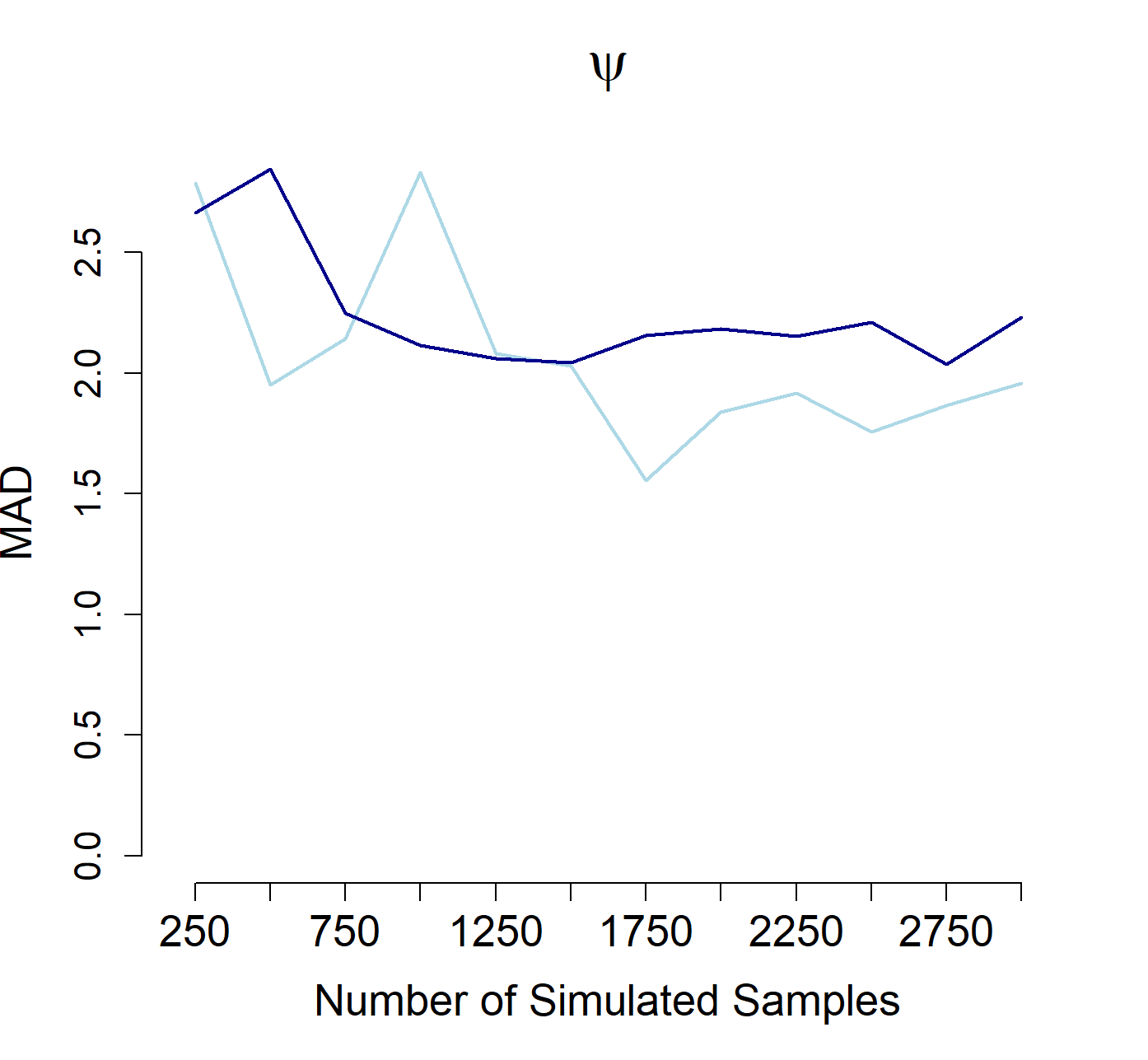}
\caption{Median absolute deviation (MAD) in the non-spatial SIR model parameters of Bayesflow and VaNBayes {\color{black}across training sample size, $N$}.}\label{f:SIR_MAD}
\end{figure}

{\color{black}Figure \ref{f:SIR_MAD} compares the two methods in terms of mean absolute deviation (MAD) of the posterior medians from the true parameter values, averaged over the 100 simulated datasets. At the lowest sample size, $N=250$, the MAD of VaNBayes is either nearly the same as Bayesflow (for estimating $\lambda$ and $\psi$), or much lower than Bayesflow, as when estimating the number of initial infected, $I_0$. As the sample size increased, the methods performed increasingly similarly.}

\subsubsection{Spatial SIR}


The spatial SIR model expands the classic SIR model to include a spatial dimension, corresponding to an additional spatial infection rate parameter that determines how neighboring locations contribute to the infection rate. We consider the model in \cite{trostle2022gaussianprocess}. At time $t$, $X_i(t)$, $Y_i(t)$ and $Z_i(t)$ are stochastic processes describing the number of susceptible, infected, and recovered people, respectively, in region $i$.  Denote the total population in region $i$ as $M_i = X_i(t)+Y_i(t)+Z_i(t)$ and the set of regions neighboring region $i $ as $\mathcal{N}_i$. Realizations of the SIR process can be approximated with a jump process which depends on the number of infected in the same region and neighboring regions, as well as the number of recovered in the same region. Denote $I_i^+(t)$ as the event that a new individual is infected at time $t$ at location $i$ and $R_i^+(t)$ as the event that an infected individual recovers at time $t$ at location $i$. Approximations for the probabilities that characterize this jump process are:
\begin{eqnarray}\label{e:spatialSIR}
    P\{I_i^+(t) | X_j(t), Y_j(t) \mbox{ for all } j\} & \approx &  \frac{\Delta t X_i(t)}{M_i}\Bigl\{\beta Y_i(s_i) + \phi\sum_{j \in \mathcal{N}_i}Y_j(t)\Bigr\}  \\
    P\{R_i^+(t) | X_i(t), Y_i(t)  \mbox{ for all } j\}  & \approx & \frac{\eta\Delta t}{M_i}Y_i(t),  
\nonumber
\end{eqnarray}
where $\Delta t$ is an arbitrarily small time frame, $\beta$ is the local infection parameter, $\phi$ is the spatial infection parameter and $\eta$ is the recovery rate.

In this simulation study, we aim to estimate the marginal posteriors of $\bgamma = (\beta, \phi, \eta)$ on a $10\times10$ grid of regions with $\sum M_i=1000$.  In Setting 1, we consider a quickly-spreading disease with parameters $\beta = 0.7, \phi = 0.8$ and $\eta = 0.5$. Setting 2 is a slower disease model with parameters $\beta = 0.5, \phi = 0.3$ and $\eta = 0.3$. To mimic a real data collection procedure, rather than assume $X$ and $Y$ are observed for all $t$, we assume they are observed at 21 time points-- the first initial time point $t_1$ and 20 evenly-spaced follow-up time points $t_j$ for $j=2,...,20$. To mimic an under-reporting of cases, the responses are Binomial-distributed random variables of the random process, $\widehat{Y}_i(t_j)\sim \mbox{Binomial}\{Y_i(t_j), p\}$ and $\widehat{Z}_i(t_j)\sim \mbox{Binomial}\{Z_i(t_j), p\}$ for all $i,j$. We will assume the under-reporting probability is $p=0.6$ and is known. At the initial time $t_1$ there are 10 infected in the (7,3) grid cell, and the rest of the population are susceptibles.

We use the prior distributions $\beta,\phi,\eta \overset{ind}{\sim} U(0.1,0.9)$. These are reasonable as uninformative priors because each of these parameters lies in $(0,1)$, and we do not expect very extreme values. The variational posterior is assumed to be the heterogeneous normal model in \eqref{e:het_normal}. We transform the parameters to the real line for use with the Gaussian model using the invertible transformation
$
\beta' = \Phi^{-1} \{ (\beta - 0.1)/0.8\},
$
where $\Phi^{-1}$ is the standard normal quantile function. After fitting the neural network on the transformed space, the posterior distribution is transformed to the original scale for presentation.

We generate $N=100,000$ synthetic datasets to train our neural networks. Since there are 4,200 responses for each dataset, we use principal component analysis (PCA) to construct summary statistics. We compute the $4,200 \times 4,200$ sample covariance matrix across the $N$ datasets and extract the leading $m$ PC scores as the summary statistics in $\bZ$. We compare $m\in\{3, 370, 950\}$ to account for 50\%, 70\% and 90\% of the variation in the model, respectively.

\begin{figure}[hbt!]
\centering 
\includegraphics[width=0.46\textwidth,page=1]{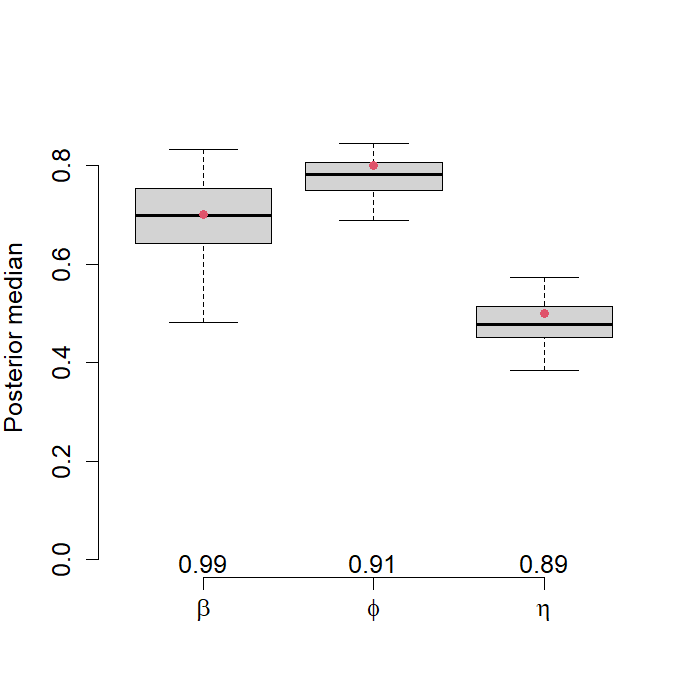}
\includegraphics[width=0.46\textwidth,page=2]{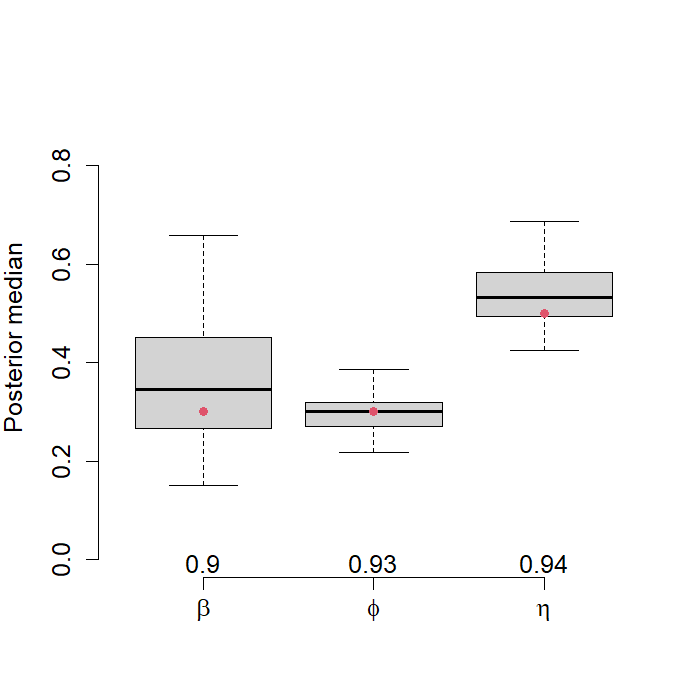}
\caption{Posterior medians of $\beta$, $\phi$, $\eta$ produced by VaNBayes for the spatial SIR model using 950 PC scores, which explain 90\% of the variation in the data. Each of the 100 points correspond to a posterior median from a simulated dataset using the true parameters. The true values of $\beta$, $\phi$, and $\eta$ are shown in red. The numbers above the $x$-axis are the coverages for the 90\% credible intervals. The figure on the left is Setting 1, and the figure on the right is Setting 2.}
\label{f:SIR_median_boxplots}
\end{figure}

After the neural networks were fitted, 100 datasets were generated from the spatial SIR model using the true parameters and fit using the trained neural networks. Table \ref{t:sir_results} compares the median absolute deviation and credible interval coverage of estimated posteriors in Settings 1 and 2 using different summary statistics.  Figure \ref{f:SIR_median_boxplots} shows how the posterior medians for each parameter are spread with respect to the true values of both settings using $m=950$ PCA scores. The posterior medians are centered around the true values and most of the 90\% credible intervals have coverage probabilities higher than 90\%. 

\begin{table}[hbt!]
\centering
 \begin{tabular}{ll|cc|cc} 
& & \multicolumn{2}{c}{Setting $1$} & \multicolumn{2}{c}{Setting $2$}\\
Parameter & PC scores & MAD  & Coverage  &MAD & Coverage \\
 \hline
$\beta$ & 3   & 0.097 & 1.00 & 0.105  & 0.99 \\
        & 370 & 0.077 & 0.97 & 0.097  & 0.94 \\
        & 950 & 0.072 & 0.99 & 0.105  & 0.90 \\
\vspace{-6pt}&&&&&\\
$\phi$ & 3   & 0.039 & 0.96 & 0.042  & 0.96\\
       & 370 & 0.036 & 0.95 & 0.026  & 0.86\\
       & 950 & 0.038 & 0.91 & 0.030 & 0.93\\
\vspace{-6pt}&&&&&\\
$\eta$ & 3   & 0.043 & 0.90 & 0.026 & 0.94 \\
       & 370 & 0.032 & 0.80 & 0.017 & 0.84\\
       & 950 & 0.041 & 0.89 & 0.059 & 0.94
\end{tabular}
 \caption{Median absolute deviation (MAD) and coverage of 90\% intervals for the spatial SIR simulation studies by parameter and the number of PC scores used as summary statistics.}
\label{t:sir_results}
\end{table}

\begin{figure}
\centering 
\includegraphics[width=0.5\textwidth,page=3]{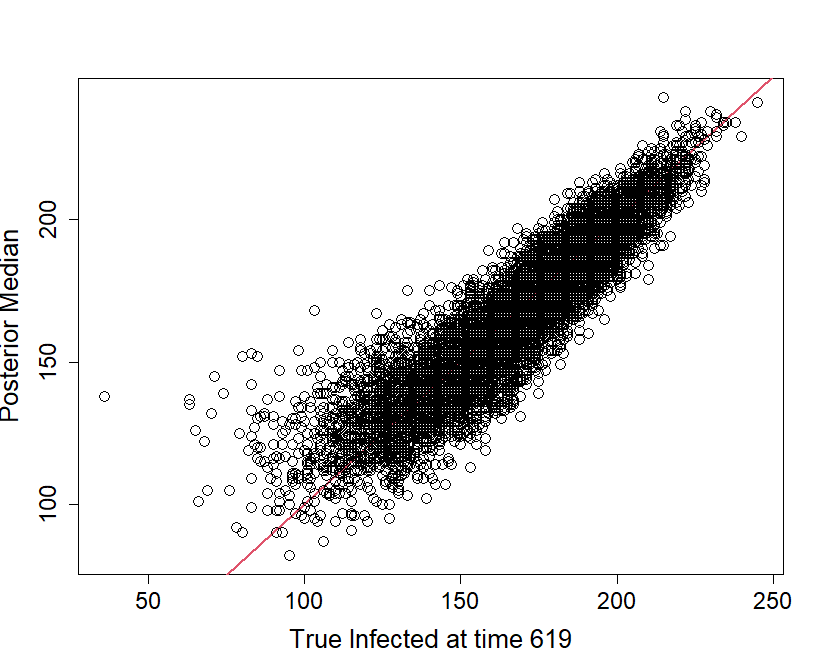}
\caption{VaNBayes posterior median of the number of infected at time $t=619$ for the spatial SIR model. The red line is the $x=y$ line.}
\label{f:spatial_SIR_scatter}
\end{figure}

{\color{black}As shown in section \ref{ss:sparse_linear_reg}, VaNBayes can be used to estimate the posteriors of discrete parameters. In this example, one potential quantity of interest would be the infected count at a particular time summing over spatial locations. Since this is a discrete quantity, it is reasonable to model it with a negative binomial distribution. To illustrate this, we implemented VaNBayes using the mean-discrepancy parametrization of the negative binomial distribution to estimate the posterior of the true counts of infected at time $t=619$. We chose this time since we found that usually there were zero infected at the end of most simulations due to the priors we chose. We used the first 950 principal component scores as predictors, exactly as described earlier. Figure \ref{f:spatial_SIR_scatter} compares the VaNBayes negative binomial posterior medians against the true count of infected at time $t=619$ across 10,000 validation simulations corresponding to the true simulation study parameters. The posterior medians exhibit a correlation of $0.91$ with the correct counts of infected at this time. The $90\%$ credible interval coverage was around $93.3\%$.}

\subsection{Spatial extremes model}

Max-stable distributions are a natural class of processes when sample maxima are observed at each site of a spatial process. They generalize the extreme value distribution to the multivariate case. 
The usefulness of these models, however, is limited by their computationally intractable likelihood function, even in moderate dimensions. Let $\{\widetilde{X}_i(\bs)\}_{\bs \in \mathcal{S}}, i = 1, \ldots, n$ be a sequence of $n$ independent replications of a continuous stochastic process in an index set $\mathcal{S}$.
If there exist sequences of continuous functions $a_n(\bs) > 0$ and  $b_n(\bs) \in \mathbb{R}$ such that
\begin{equation}
    X(\bs) = \lim_{n\to\infty} \frac{\max^n_{i=1} \widetilde{X}_i(\bs) - b_n(\bs) }{a_n(\bs)}, \quad {\bs \in \mathcal{S}},
    \label{eq:maxstab}
\end{equation}
then the limit process $X(\bs)$ is a max-stable process \citep{de1984spectral}. The one-dimensional marginal distributions are from the class of generalized extreme value (GEV) distributions, denoted $Y \sim \mbox{GEV}(\mu, \sigma, \xi)$, with distribution function
\begin{equation}
    F(y; \mu, \sigma, \xi) = \mbox{exp} \left[ - \Biggl\{ 1 + \frac{\xi (y - \mu)}{\sigma} \Biggl\} ^{-1/\xi}_{+} \right], \quad \mu \in \mathbb{R}, \quad \sigma > 0, \quad \xi \in \mathbb{R},
    \label{eq:gev}
\end{equation}
where $a_{+} = \mbox{max}(0; a)$ and $\mu, \sigma$ and $\xi$ are respectively location, scale and shape parameters \citep{haan2006extreme}. 
Following \citet{de1984spectral}, a max-stable process can be constructed by its spectral characterization. Different forms of spatial dependence can be constructed depending on the choice of the stochastic process in this representation. In what follows, we consider the Brown-Resnick process \citep{kabluchko2009stationary}, a widely used parametric form in spatial extremes with the spatial dependence described by the semivariogram $\gamma(\bh) = (\lVert\bh\rVert/\lambda)^\nu$, where $\bh$ is the spatial separation distance, and range $\lambda > 0$ and smoothness $\nu \in (0, 2]$ are parameters.

The parameters of interest are the ones from the marginal GEV and the max-stable spatial process,  $\bgamma = \btheta=[\log(\lambda), \log\{ \nu/(2 - \nu) \}, \mu, \log(\sigma), \xi ]$.
We simulate 100 non-gridded spatial locations uniformly and independently on $[0,10]$.   
To train the neural network, we generate $N=10,000$ samples from prior distributions 
$\theta_1, \theta_2, \theta_3, \theta_4 \sim \mbox{Normal}(0,1)$ and $\theta_5 \sim \mbox{Normal}(0,0.1)$.
The summary statistic for describing the spatial structure is the extremal coefficient, $\omega(\bs_1; \bs_2) \in [1, 2]$, where  $\omega(\bs_1; \bs_2) = 1$ corresponds to perfect dependence and $\omega(\bs_1; \bs_2) = 2$ to independence  \citep{cooley2006variograms}.  
These summary statistics are computed as averages of the empirical extremal coefficient over 10 equally-spaced bins using 50 data replicates. 
We also use the empirical quantiles $(q_{0.5}, q_{0.7}, q_{0.9}, q_{0.95}, q_{0.99}, q_{1})$ from data at the 100 locations and 50 replicates as summary statistics to capture the marginal data distribution. 
The parameters are modeled independently using the heterogeneous normal model in \eqref{e:het_normal}. 
The neural network architecture and tuning parameters are the same as in the previous examples described in Sections~\ref{ss:sparse_linear_reg} and \ref{ss:autologistic}.

\begin{figure}
\centering 
\includegraphics[width=0.6\textwidth,page=1]{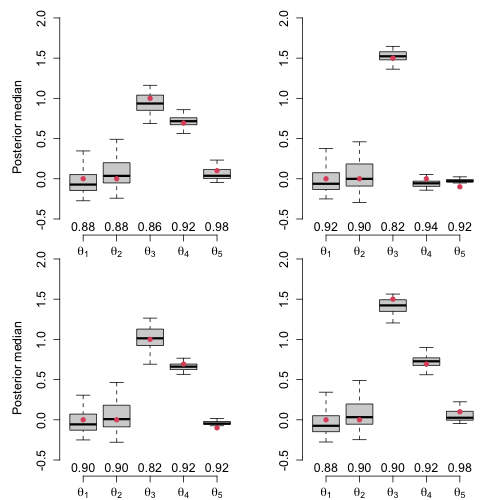}
\caption{Sampling distribution of the posterior median of the coefficients for the spatial extremes model. The panels differ by the true values, which are shown as red dots and the coverage percentages of 90\% posterior intervals are given below the boxplots. 
\label{f:maxstab-boxplot}}
\end{figure}

\begin{figure}
\centering 
\includegraphics[width=0.5\textwidth,page=1]{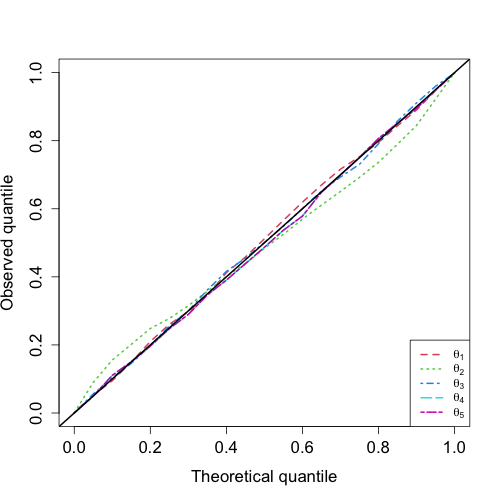}
\caption{QQ-plot of the probability integral transform for the estimated parameters for the spatial extreme model. 
\label{f:maxstab-scatter}}
\end{figure}

We test the performance of the trained neural network with different true parameter value scenarios. 
Figure~\ref{f:maxstab-boxplot} displays boxplots of the posterior medians from 50 datasets at four different scenarios of true parameter values (red dots): $\btheta = \{0, 1.1, 1, 0.7, -0.1\}$, $\btheta = \{0, 0, 1.5, 0.7, -0.1\}$, $\btheta = \{0.7, 1.1, 1, 0.7, 0.1\}$ and $\btheta = \{0, 1.1, 1.5, 0.7, 0.1\}$ (from left to right and top to bottom).
The coverage percentages of 90\% credible intervals are given below the boxplots. This shows that the parameters are always estimated well, with the uncertainty in the estimation differing mostly by parameter type than by scenario.
Figure~\ref{f:maxstab-scatter} shows that the parameter inferences are calibrated, with slightly too conservative estimations for $\theta_2$. 

\section{Spatiotemporal analysis of the Zika virus in Brazil}\label{s:data analysis}

We use VaNBayes to estimate the spread of the Brazilian Zika virus using the records found in \cite{trostle2022gaussianprocess}. The data consist of weekly counts of the number of new cases by state for the 40-week period from the 41st week of 2015 to the 28th week of 2016. Code for cleaning this data can be found at: https://github.com/jptrostle/SpatialSIRGPMC. The Zika virus is particularly dangerous to unborn children of women who have contacted the Zika virus, resulting in severe birth defects and neurodevelopment impairment \citep{MARBANCASTRO2021162}, so  modeling and forecasting its spread is potentially valuable, but computationally challenging, as outlined in Section \ref{s:SIR}. 

\subsection{Spatial SIR model}

We use the model proposed by  \cite{trostle2022gaussianprocess}.  The true disease status follows the model in Section \ref{s:SIR}, with the exception that the local infection rate ($\beta$ in (\ref{e:spatialSIR})) in state $i$ is $\exp(\beta_0+X_i\beta_1)$, where $X_i$ is the log population density in state $i$.  Given the true disease status, the observed number of new cases in state $i$ and week $t$ is assumed to follow a negative binomial distribution with mean $p_i\{Y_i(t)-Y_i(t-1)\}$ and over-dispersion parameter $\nu$, where $p_i$ is the reporting rate for state $i$. The three parameters of interest are $\bgamma = (\beta_0, \beta_1, \phi)$, where $\phi$ is the spatial infection rate.  The recovery rate $\eta$, overdispersion parameter $\nu$, and reporting rates $p_i$ for each state $i=1,...,27$ are nuisance parameters. $\eta$ and all $p_i$ are set to the values in \cite{trostle2022gaussianprocess} and  uninformative priors are used for the remaining parameters: $\beta_0 \sim \text{Uniform}(-3, 1)$, $\beta_1 \sim \text{Uniform}(-1,1)$, $\log(\phi) \sim \text{Normal}(-2,1)$ and $\nu\sim\text{Uniform}(1.01,10)$. 

\subsection{Model Fitting and Results}

The \texttt{SimInf} package \citep{SimInf} was used to generate realizations from the spatial SIR model.  We train the neural network using 80,000 datasets drawn from the model with the prior as the training distribution. After generating our data, we used VaNBayes to estimate the posterior of $\beta_0$, $\beta_1$, and $\phi$; we use different combinations of three neural network architectures and four sets of summary statistics to tune the posterior approximation. All neural networks have two fully connected hidden layers. The smallest neural network has 50 nodes in the first hidden layer and 10 nodes in the second hidden layer, denoted (50, 10). The medium and large neural networks have (100, 25) and (150, 50) nodes, respectively. As summary statistics, we consider 4, 6, 9, and 20 principal components (PCs) of the infected counts of the 27 states and 40 weeks across the 80,000 simulated datasets, which explain 80\%, 90\%, 95\%, and 99\% of the variance across the simulated datasets respectively. 



\begin{figure}[hbt!]
\centering 
\includegraphics[width=0.46\textwidth,page=1]{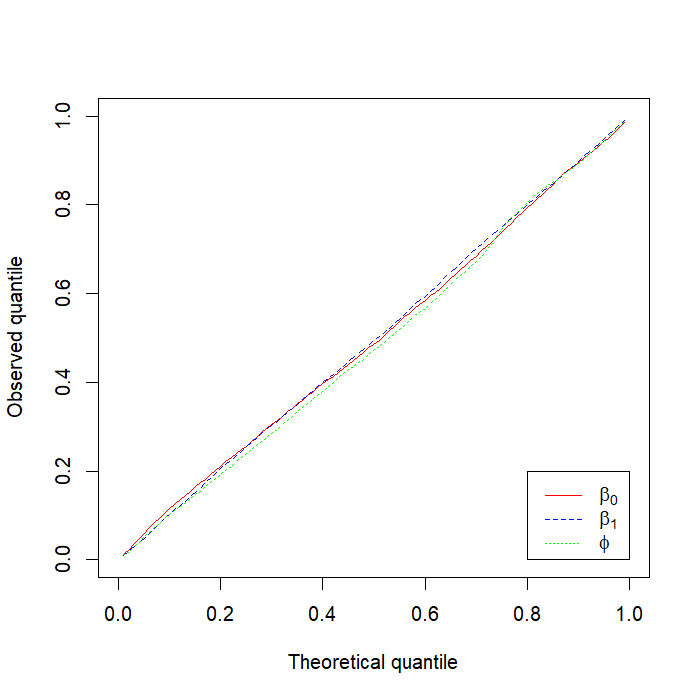}
\caption{The PIT plot for the 20 principal components and large neural network VaNBayes configuration for the spatial SIR model for the Zika virus. \label{f:RDA_PIT}}
\end{figure}

Table \ref{t:log_scores} shows the variational posterior's log-scores for $\beta_0$, $\beta_1$, and $\phi$ across the different VaNBayes configurations. The model with the best fit is the largest neural network with the 20 principal components. Generally, the larger neural networks fit the data better than the smaller neural networks and the models with more principal components fit the data better than those with fewer principal components. The PIT plot in Figure \ref{f:RDA_PIT} shows that this VaNBayes configuration seems to fit the data well. 


\begin{table}
\centering
 \begin{tabular}{ll|ccc} 
&& \multicolumn{3}{c}{Number of neural network nodes $(L_1,L_2)$}  \\
Parameter & Model  & $(50, 10)$  & $(100, 25)$ & $(150, 50)$   \\
\hline
$\beta_0$ 
& 4 PCs  & $-1.339$ & $-1.303$ & $-1.288$ \\  
& 6 PCs  & $-1.296$ & $-1.313$ & $-1.283$ \\  
& 9 PCs  & $-1.265$ & $-1.219$ & $-1.200$ \\  
& 20 PCs & $-1.179$ & $-1.166$ & {\bf $-1.136$} \\  
& \vspace{-6pt} \\
$\beta_1$ 
& 4 PCs  & $-1.360$ & $-1.340$ & $-1.341$ \\  
& 6 PCs  & $-1.328$ & $-1.351$ & $-1.354$ \\  
& 9 PCs  & $-1.306$ & $-1.292$ & $-1.257$ \\  
& 20 PCs & $-1.240$ & $-1.191$ & {\bf $-1.177$} \\  
& \vspace{-6pt} \\
$\phi$ 
& 4 PCs  & $-1.327$ & $-1.338$ & $-1.326$ \\  
& 6 PCs  & $-1.279$ & $-1.331$ & $-1.304$ \\  
& 9 PCs  & $-1.224$ & $-1.240$ & $-1.214$ \\  
& 20 PCs & $-1.144$ & {\bf $-1.120$} & $-1.122$ \\  
\end{tabular}
\caption{The average log-scores of the variational posterior for each parameter of interest across different configurations of VaNBayes. Larger values of the log-scores implies the variational posterior fits the data better. }
\label{t:log_scores}
\end{table}

\begin{figure}[hbt!]
    \centering
    \begin{minipage}{1\textwidth}
        \centering
        \includegraphics[width=0.46\textwidth,page=1]{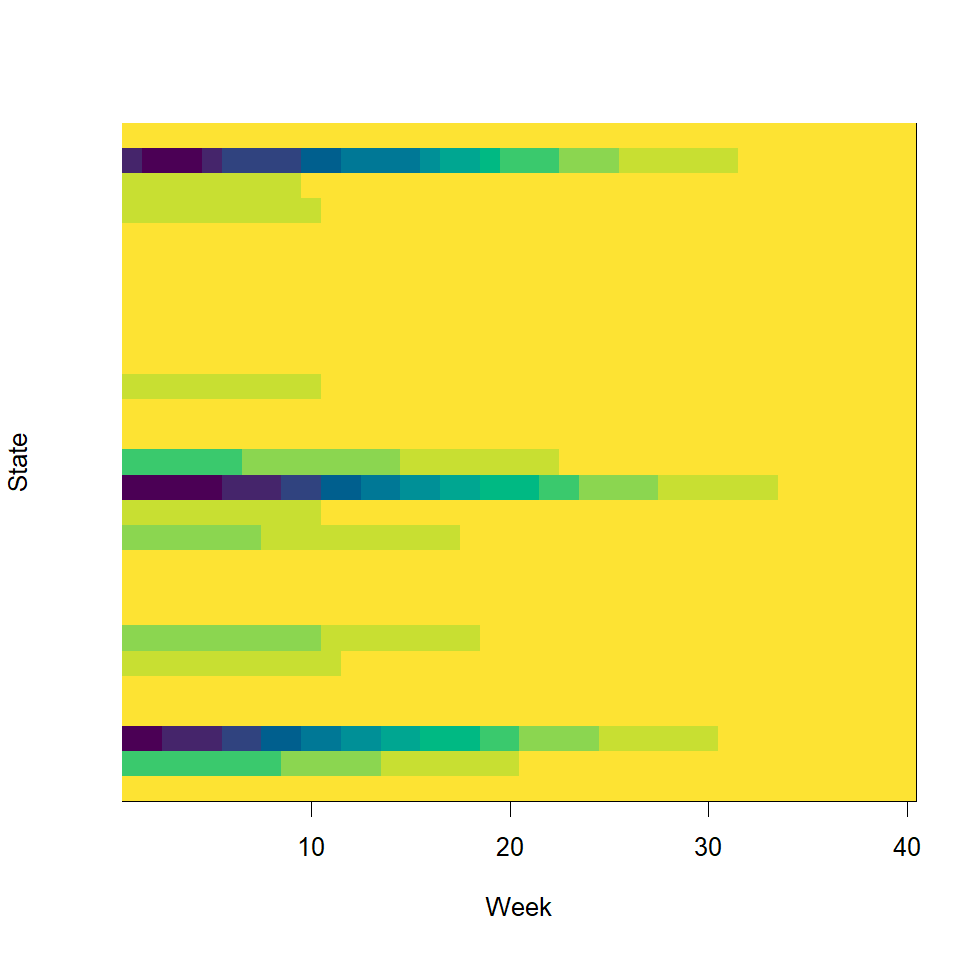}
        \includegraphics[width=0.46\textwidth,page=2]{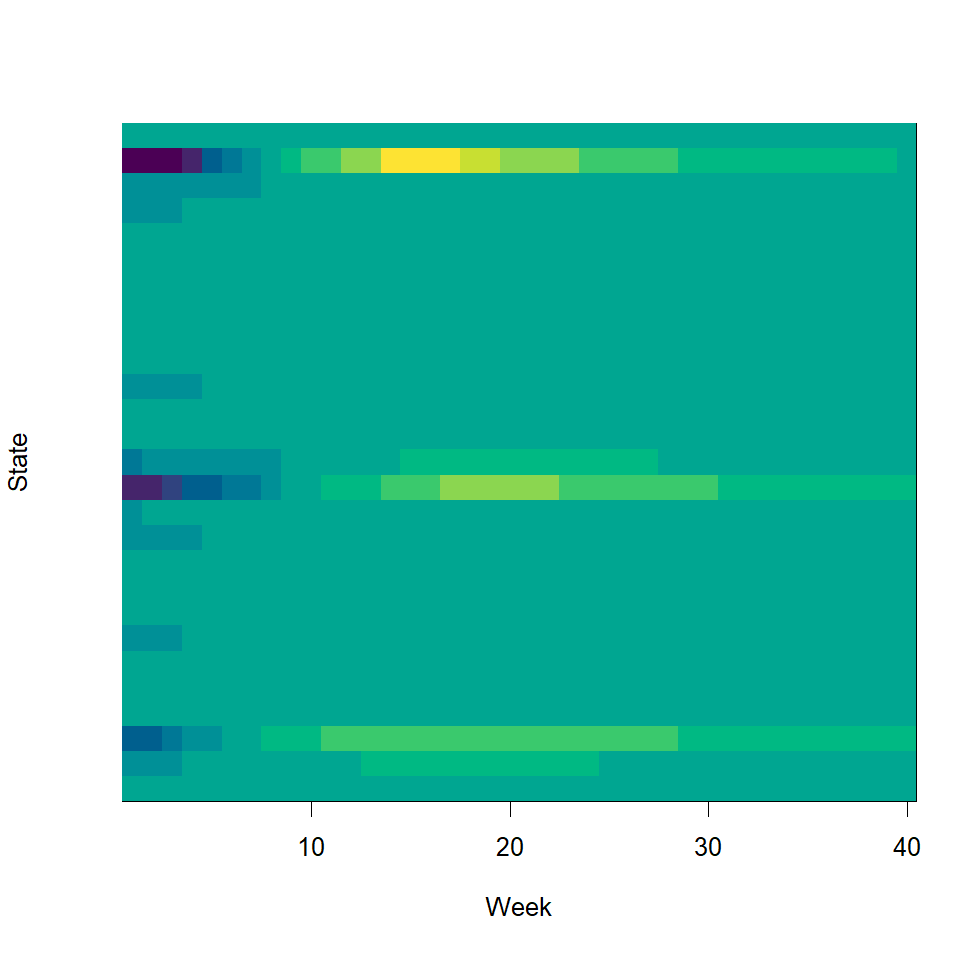}
            \end{minipage}\vfill
    \begin{minipage}{.99\textwidth}
        \centering
        \includegraphics[width=0.47\textwidth,page=1]{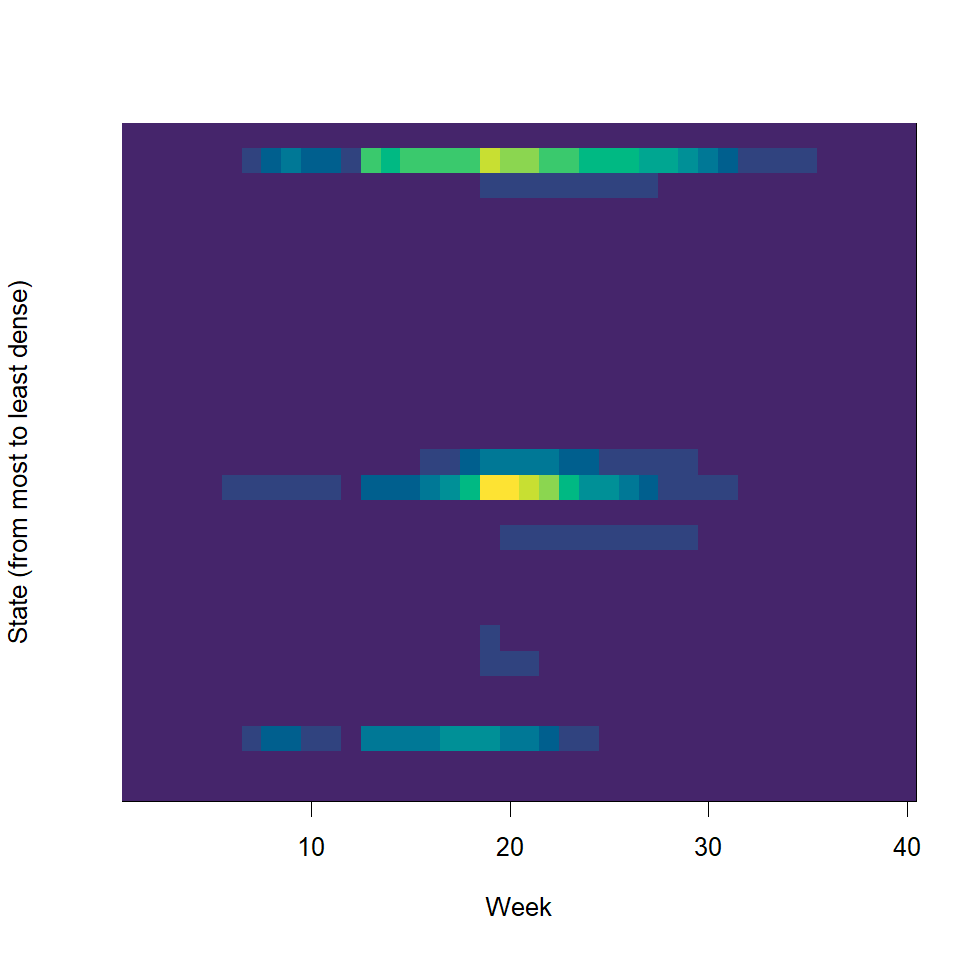}
        \includegraphics[width=0.47\textwidth,page=2]{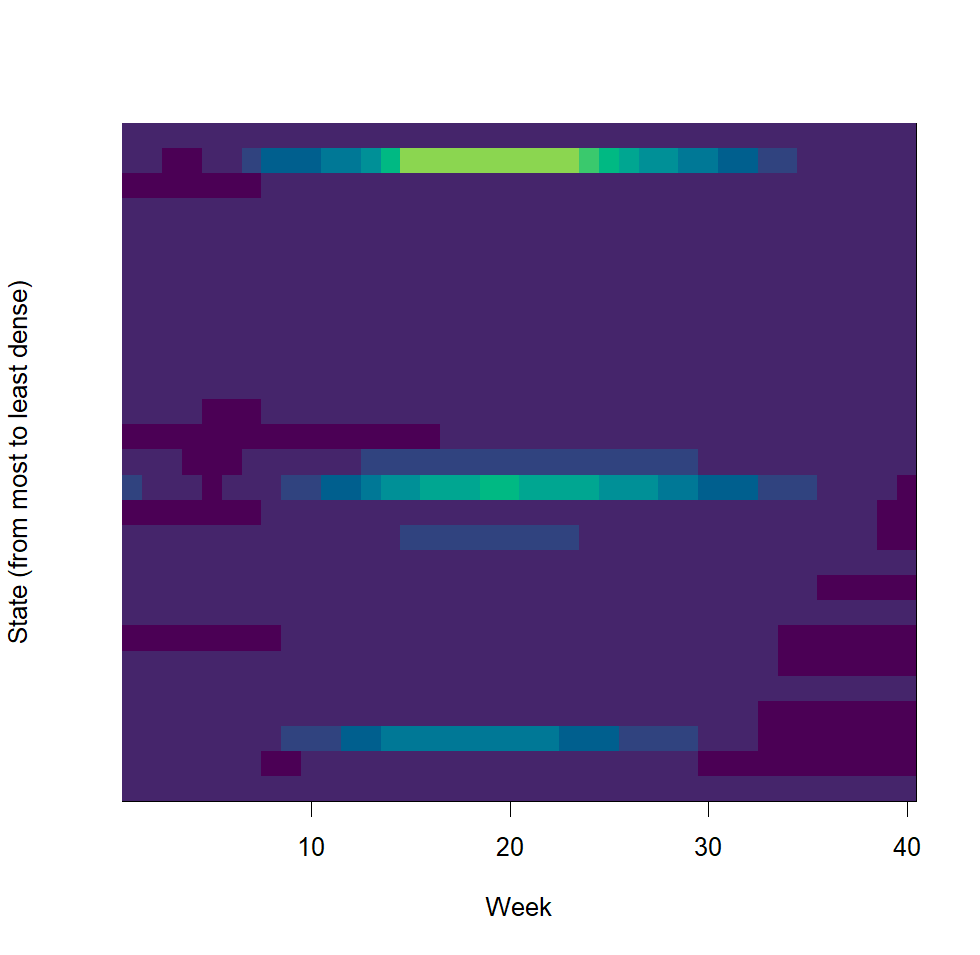}
        \end{minipage}
    \caption{(Top Row) The loading values of the first and second PC's, respectively. The rows of each heatmap correspond to different Brazilian states, starting with the most densely populated at the top and least densely populated on the last row. (Bottom Row) The infected counts for the data and the reconstructed data using the 20 PC's.\label{f:real_data_heatmaps}}
\end{figure}

We explore the contribution of data from different Brazilian states to the posteriors by examining the PC factor loadings.
The top row of Figure \ref{f:real_data_heatmaps} arranges the loading matrix of the $27\times 40$ data points into a heatmap of states versus weekly time. High levels of the first PC (top left) capture change in the infected counts of three states within the first 20 weeks. The second PC (top right) captures change in the infected counts in those states around weeks 15-20. The bottom row of Figure \ref{f:real_data_heatmaps} shows that 20 PC's can reconstruct the observed data fairly well. 
The three Brazilian states identified in the PCA loading heatmaps are Rio de Janeiro (highest prominent row), Bahia (middle prominent row), and Mato Grosso (lowest prominent row). 
The identified states are the second to fourth most populous in Brazil. Sao Paulo is the most populous state, but it borders Rio de Janeiro, which has a higher population density.

\begin{figure}[hbt!]
\centering 
\includegraphics[width=0.32\textwidth,page=1]{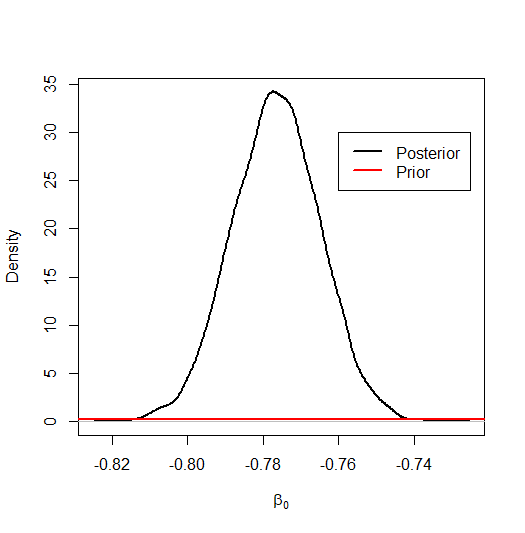}
\includegraphics[width=0.32\textwidth,page=2]{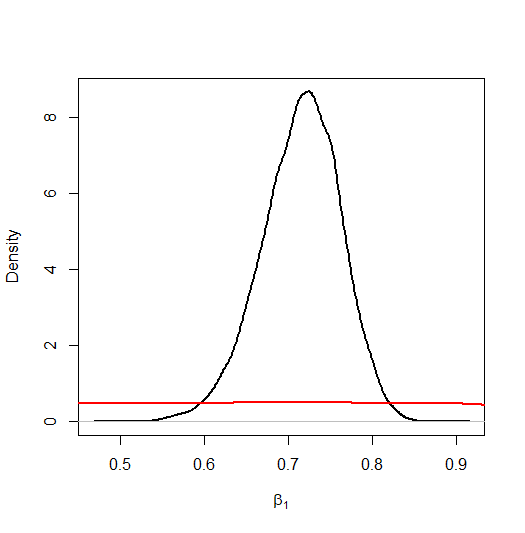}
\includegraphics[width=0.32\textwidth,page=2]{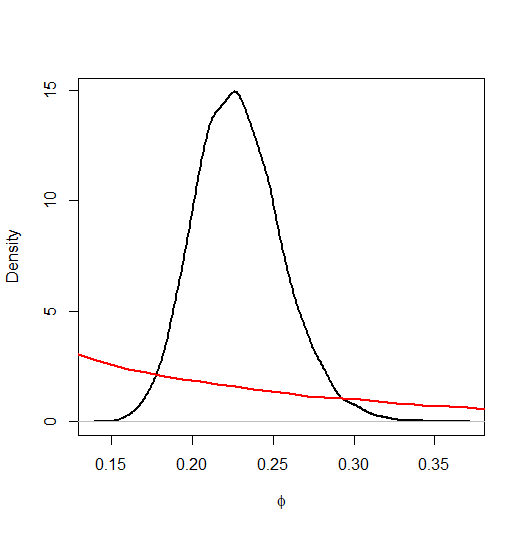}
\caption{Prior and posterior distributions for the parameters in the SIR model for the Zika data. \label{f:prior_posterior_comparisons}}
\end{figure}

Figure \ref{f:prior_posterior_comparisons} shows the variational posteriors for each variable alongside the priors. The 95\% credible intervals are $(-0.80, -0.75)$, $(0.61, 0.80)$, and $(0.18, 0.29)$, for $\beta_0$, $\beta_1$, and $\phi$ respectively. The credible intervals imply that the $\beta_0$ parameter is negative, while the $\beta_1$ parameter is positive. This agrees with our intuition, as one would expect the population density to be proportional to the local infection rate. These findings generally agree with the data analysis of \cite{trostle2022gaussianprocess}.

\section{Discussion}\label{s:discussion}

In this paper, {\color{black}we demonstrate the utility of variational approximation for simulation-based inference applied to challenging environmental problems}. We formalize {\color{black}this approach}, give advice on choosing the proposal distribution and variational posterior approximation, and show that our estimated posterior distribution converges to the true posterior distribution with respect to the Kullback-Leibler divergence under a few assumptions. We also demonstrate that VaNBayes works well in a large variety of settings, notably a spatial SIR model and a max-stable process model. {\color{black}We compare it to Bayesflow and show that carefully-chosen parametric assumptions can improve fit for high-dimensional problems and small training data, and facilitate approximations to discrete posterior distributions.}


{\color{black}As with all parametric approaches, VaNbayes works well when the variational assumption is the correct choice; while model diagnostics could give insight as to how to better choose the variational distribution, doing so may take time. Furthermore, this paper assumes that we are only interested in low-dimensional summaries of the posterior distribution. However, in some cases, one may be interested in the joint distribution of the parameters of interest, which is challenging to fit using VaNBayes.}


While the choice of the summary $\bZ$ is context dependent, its selection can be formulated as a statistical modeling problem, which is especially feasible when we have a large amount of simulated data. Moreover, it is possible to select $\bZ=\bY$ when no suitable summary statistic can be found.

Our approach to finding posterior estimates of parameters in models with intractable likelihoods circumvents the computational and modeling costs of traditional Bayesian and simulation-based Bayesian estimation methods in highly complex models from which it is easy to simulate. As models increase in complexity to adapt to ever more challenging statistical problems, we anticipate simulation-based methods like VaNBayes will become increasingly popular for their ability to flexibly and reliably provide Bayesian inference.


\appendix
\section{Proof for Theorem 1}
\label{s:A1}

Assume that $\bY$ and $\bZ$ are continuous; a similar argument holds in the discrete case and is omitted. We first show that the estimators in \eqref{e:wMLE} converge in probability to the maximizers of the expectation for the variational posterior of $\bgamma$ with respect to the prior distribution $\pi(\btheta)$. The expectation is given by
\begin{align*}
o(\bW)&=\E_{\pi(\boldsymbol{\theta})} ( \log [p \{ \bgamma | a(\bZ;\bW) \}  ] )\\
&= \int \pi(\btheta)\log \left[p \left\{ \bgamma | a(\bZ;\bW)\right\}  \right]  d\btheta \\
&= \int \pi(\btheta)\frac{\Pi(\btheta)}{\Pi(\btheta)}\log \left[p \left\{ \bgamma | a(\bZ;\bW) \right\}  \right]  d\btheta \\
&=\E_{\Pi(\boldsymbol{\theta})} \left( \frac{\pi(\btheta)}{\Pi(\btheta)} \log \left[p \left\{ \bgamma | a(\bZ;\bW) \right\} \right] \right).
\end{align*}

Treating $(\btheta_i , \bZ_i)$ as observed data pairs, the estimators in \eqref{e:wMLE} maximize an empirical version of this expectation, given by
\begin{align*}
O(\bW)&=\sum \limits_{i=1}^N \frac{\pi(\btheta_i)}{\Pi(\btheta_i)} \log \left[p \left\{ \bgamma_i | a(\bZ_i,\bW) \right\} \right].
\end{align*}
By the weak law of large numbers, $O(\bW) \stackrel{p}{\rightarrow} o(\bW)$ as $N \rightarrow \infty$. By smoothness of $O(\cdot)$ and $o(\cdot)$, the maximum of the former converges in probability to the maximum of the latter. Defining $\widehat{\bW}$ to be the argument maximum of $O(\bW)$, 
\begin{equation*}
\sum \limits_{i=1}^N \frac{\pi(\btheta_i)}{\Pi(\btheta_i)} \log \left[p \left\{ \bgamma_i | a(\bZ_i;\widehat{\bW})\right\} \right] 
\stackrel{p}{\rightarrow} \max \limits_{\bW} \E_{\pi(\boldsymbol{\theta})} \left( \log \left[p \left\{ \bgamma | a(\bZ;\bW)\right\}  \right] \right).
\end{equation*}
Although this is an asymptotic result, we simulate $N$ data pairs prior to fitting the neural networks. Hence, for arbitrarily large $N$, our objective function does not rely on $\Pi(\btheta)$, and instead only relies on the term $\E_{\pi(\boldsymbol{\theta})} \left( \log \left[p \left\{ \bgamma | a(\bZ;\bW) \right\}  \right] \right)$.

\section*{Acknowledgements}

This work was partially supported by National Science Foundation grants DMS2152887 and DMS2342344 and a grant from North Carolina State University's Office of Global Engagement. The authors declare no conflicts of interest.

\bibliography{refs}

\end{document}

%% file: commands.tex
\newcommand{\E}{\mathbb{E}}

\newcommand{\btheta}{ \mbox{\boldmath $\theta$}}

\newcommand{\bbeta}{ \mbox{\boldmath $\beta$}}

\newcommand{\bgamma}{ \mbox{\boldmath $\gamma$}}

\newcommand{\bnu}{ \mbox{\boldmath $\nu$}}

\newcommand{\bX}{ \mbox{\bf X}}

\newcommand{\bZ}{ \mbox{\bf Z}}

\newcommand{\bY}{ \mbox{\bf Y}}

\newcommand{\bh}{ \mbox{\bf h}}

\newcommand{\bs}{ \mbox{\bf s}}

\newcommand{\bu}{ \mbox{\bf u}}
\newcommand{\bv}{ \mbox{\bf v}}

\newcommand{\bW}{ \mbox{\bf W}}
\newcommand{\bw}{ \mbox{\bf w}}

\newcommand{\iid}{\stackrel{iid}{\sim}}
\newcommand{\indep}{\stackrel{indep}{\sim}}

\newcommand{\beq}{ \begin{equation}}
\newcommand{\eeq}{ \end{equation}}
\newcommand{\beqn}{ \begin{eqnarray}}
\newcommand{\eeqn}{ \end{eqnarray}}

\usepackage{caption}